\documentstyle[preprint,flushrt]{aastex}
\begin{document}
\newcommand{\redshift} {\left( \frac{1+z}{10} \right)}
\newcommand{\SFR} {{\rm \left( \frac{SFR}{1 M_{\odot} \, yr^{-1}}
\right)}}  
\newcommand{\Bfield} {{\rm \left( \frac{B}{10 \mu G} \right) }}
\newcommand{\efficiency} {\left( \frac{\epsilon}{0.1} \right)}
\newcommand{\freq} { \left( \frac{\nu}{1 \, {\rm GHz}} \right)}
\newcommand{\radrel} { {\rm min} \left[ \redshift^{-4},
\left( \frac{U_{\gamma}}{ 4.2 \times 10^{-9} \, {\rm erg \, cm^{-3}}}
\right)^{-1} \right] }
\newcommand{\density} {\left( {\rm \frac{{\rm n}}{1 \, {\rm cm^{-3}}}}
\right)}
\newcommand{\sfryr}{{\rm M_{\odot} \, yr^{-1}}}

\title{Reionization by Hard Photons: I. \\
X-rays from the First Star Clusters}
\author{S. Peng Oh\altaffilmark{1}\\
Princeton University Observatory, Princeton, NJ 08544;
peng@astro.princeton.edu}
\altaffiltext{1}{Current address: Theoretical Astrophysics, Mail Code
130-33, California Institute of Technology, Pasadena, CA 91125; peng@tapir.caltech.edu}
\begin{abstract}
Observations of the Ly$\alpha$ forest at $z \sim 3$ reveal an average
metallicity ${\rm Z} \sim 10^{-2} Z_{\odot}$. The high-redshift supernovae
that polluted the IGM also accelerated relativistic electrons. Since
the energy density of the CMB $\propto (1+z)^{4}$, at high redshift
these electrons cool via inverse Compton scattering. Thus, the first
star clusters emit X-rays. Unlike stellar UV ionizing photons, these
X-rays can escape easily from their host galaxies. This has a number
of important physical consequences: (i) Due to their large
mean free path, these X-rays can quickly establish a universal
ionizing background and partially reionize the universe in a
gradual, homogeneous fashion. If X-rays formed the dominant ionizing background, the universe would have more closely resembled
a single-phase medium, rather than a two-phase medium. (ii) X-rays can reheat the universe to higher temperatures than possible with UV
radiation. (iii) X-rays counter the tendency of UV
radiation to photo-dissociate ${\rm H_{2}}$, an important coolant in the
early universe, by promoting gas phase ${\rm H_{2}}$ formation. The
X-ray production efficiency is calibrated to local observations of
starburst galaxies, which imply that $\sim 10 \%$ of the supernova
energy is converted to X-rays. While
direct detection of sources in X-ray emission is difficult, the
presence of relativistic electrons at high redshift and thus a minimal
level of X-ray emission may be inferred by synchrotron emission
observations with the Square Kilometer Array. These sources may constitute a significant fraction of the unresolved hard X-ray background, and can account for both the shape and amplitude of the gamma-ray background. This paper discusses the existence and observability of
high-redshift X-ray sources, while a companion paper 
models the detailed reionization physics and chemistry.  
\end{abstract}
%%%%%\clearpage

\section{Introduction}

While the theoretical literature on the epoch of reionization is large
and increasing rapidly, our empirical knowledge of this period in the
history of the universe is scant and may be succintly summarized: (i) The universe is likely to have been reionized in the period
$5.8 < z_{r} < 35$, where the lower bound arises from the lack of
Gunn-Peterson absorption in the spectra of high redshift quasars
(Fan et al 2000), and the upper bound comes from the observation
of small scale power in Cosmic Microwave Background anistropies (Griffiths et al 1999). In CDM
cosmologies, non-linear objects above the cosmological Jeans mass
$10^{4-5} M_{\odot}$
first collapse during this period. (ii) The presence of metals in the
Ly$\alpha$ forest implies that significant star formation took place
at high redshift (Songailia \& Cowie 1996, Songailia 1997). (iii) COBE constraints on the Compton
y-distortion of the CMB (Wright et al, 1994, Fixsen et al 1996) implies that the IGM was
not heated to high temperatures. This means it is unlikely that the
hydrogen and helium in the IGM were collisionally reionized.  

Thus, the current state of observations is consistent with scenarios
in which the universe was reionized by an early generation of stars or
quasars. While we do not know whether stars or quasars were the
dominant source of ionizing photons, the observational and theoretical
case for quasars is somewhat more uncertain. Extrapolation of empirical quasar luminosity functions to
high redshift do not yield enough ionizing photons to maintain the
observed lack of Gunn-Peterson absorption at $z \sim 5$ (Madau, Haardt \& Rees 1999). The
requisite steepening of the faint end slope at high redshift necessary to boost ionizing photon production is
constrained by the lack of red, point-like sources in the Hubble Deep
Field (Haiman, Madau \& Loeb 1999); the authors find that AGN
formation must have been suppressed in halos with $v_{c} < 50-75 \,
{\rm km \, s^{-1}}$. We do not have a sufficiently firm
understanding of the formation and fueling of supermassive black holes
to assert on theoretical grounds that AGNs must have been present at high
redshift. On the other hand, a minimal level of high-redshift star
formation is guaranteed by the observed metal pollution of the IGM. 

Theoretical scenarios in which stars or quasars figure predominantly
have been calculated in detail. Our ignorance of the efficiency of gas
fragmentation, and star/black hole formation as a function
of halo mass, make the prediction of observable differences between
these two scenarios very uncertain. Indeed, if
one normalizes assumed emissivities to a fixed reionization
epoch, differences between the two scenarios boil down to: (i) stars result in
supernovae, which inject dust, metals, and entropy into the host
galaxy and surrounding IGM, which affects subsequent chemistry and
cooling, (ii) quasars have a significantly harder spectrum than
stars. In particular, they produce X-rays. 

In this paper, I emphasize a hitherto neglected
fact: high redshift supernova also produce
X-rays, both by thermal emission from the hot supernova remnant, and
inverse Compton scattering of soft photons by relativistic electrons
accelerated by the supernova. Considerable X-ray emission is already
observed in starburst galaxies at low redshift (e.g., Rephaeli et al
1995), and the efficiency of most proposed X-ray production mechanisms should {\it
increase} with redshift (e.g., explosions take place in a denser
medium at high redshift, hardening expected thermal emission; inverse Compton
scattering becomes more efficient since the CMB provides a ready
supply of soft photons $U_{CMB} \propto (1+z)^{4}$). Thus, the SED of
high-redshift star forming regions is considerably harder than has
been previously assumed. This blurs the distinction between stellar/quasar
reionization scenarios, and has a number of important physical
consequences:
\begin{itemize}
{\item \bf Escape fraction} The escape fraction of UV ionizing photons in the
local universe is small, $\sim 3-6 \%$ (Leitherer et al 1995,
Bland-Hawthorne \& Maloney 1999, Dove, Shull \& Ferrara 2000) and is
expected to decrease with redshift (Wood \& Loeb 1999, Ricotti \&
Shull 1999). On the other hand, X-rays can
escape freely from the host galaxy. Thus, processing by the host ISM
may imply that the universe was reionized by a significantly harder
spectrum than previously assumed.    

{\item \bf Reionization topology} Photons from stellar spectra have a
short mean free path and thus a sharply defined ionization front. This
fact gives rise to the conventional picture of expanding HII bubbles
embedded in the neutral IGM. The spectra of quasars is significantly
harder and exert an influence over a larger distance (which is why it
is much more difficult to perform numerical simulations of reionization by quasars; see Gnedin
1999). Nonetheless, for quasars, $\nu L_{\nu} \propto \nu^{-0.8}$ (Zheng et al 1997, although note that the observations were only in the radio-quiet AGN subsample at energies up to 2.6 Ry) and
most of the energy for ionization lies just above the Lyman edge. Thus, there is still a sharply defined
HII region and a thin ionization front where the ionization fraction
drops sharply. By contrast, for the inverse Compton case $\nu L_{\nu}
\sim {\rm const}$, there is equal power per logarithmic interval, and thus
there is no preferred energy scale. In particular, there is no
preferred scale for the mean free path of ionizing photons. When the
universe is largely neutral, it is optically thick even to hard
photons and all photons with energies $E < E_{thick}=1.5
\redshift^{0.5} x_{HI}^{1/3} \, {\rm keV}$ (where $x_{HI}$ is the mean
neutral fraction) are absorbed across a
Hubble volume. While the (more numerous) soft photons can only travel a short
distance before ionizing neutral HI and HeI, the (less
numerous but more energetic) hard photons will be able to travel
further and ionize an equivalent number of photons by
secondary ionizations. Thus, even
if sources are distributed very inhomogeneously, reionization will be
a fairly homogeneous event, with a largely uniform ionizing background and fluctuations in ionization fraction determined mainly by gas clumping. Instead of an two-phase medium in whose HII filling
fraction increases with time, the early IGM may have been a single
phase medium whose ionization fraction increases with time.  

{\item  \bf Increased reheating} A hard spectrum can reheat the IGM to
considerably higher temperatures than soft stellar spectra, both
through photoionization heating and Compton heating. A soft spectrum loses thermal contact with the IGM once HI and HeI are completely ionized (at the mean IGM density, HI has a recombination
time longer than the Hubble time for $z<10$), and the gas cools adiabatically due to the expansion of the universe (Hui \& Gnedin 1997). By contrast, a hard spectrum can continually transfer large amounts of energy from
the radiation field to the IGM by ionizing HeII, which recombines
rapidly (Miralda-Escude \& Rees 1994). This feedback
mechanism is important in increasing the Jeans mass, a proposed
mechanism for preventing excessive cooling and star formation at high redshift (e.g., Prunet \& Blanchard 1999). The higher IGM temperatures may also explain why
observed Ly$\alpha$ forest line widths are commonly in excess of that
predicted by numerical simulations (Theuns et al 1999, Ricotti et al 2000). 

{\item \bf Early universe chemistry} ${\rm H_{2}}$ is an extremely important
coolant in the metal-free early universe. While the neutral IGM is
optically thick to UV ionizing photons, it is optically thin to
photons longward of the Lyman limit (except at wavelengths
corresponding to higher order hydrogen Lyman resonance lines, as well
as $H_{2}$ resonance lines). In particular, photons in the 11.2-13.6 eV range quickly establish a soft UV background which
photodissociates ${\rm H}_{2}$ via the Solomon process, shutting down
subsequent star formation (Haiman, Rees \& Loeb 1997, Cicardi, Ferrara
\& Abel 1998), unless the $H_{2}$ opacity is sufficient to reduce the
photo-dissociation rate (Ricotti, Gnedin \& Shull 2000). If X-rays are present in the early universe, they can
counter this $H_{2}$ destruction. The IGM is also optically thin to X-rays, which
can penetrate dense clouds of gas and promote gas phase ${\rm H_{2}}$
formation ${\rm H + e^- \rightarrow H^- + \gamma}$ and ${\rm H^- + H
\rightarrow H_2 + e^-}$ by increasing the abundance of free
electrons. Haiman, Abel \& Rees (1999) show that if quasars were the
dominant ionizing sources in the early universe, gas cooling and thus star formation can
continue unabated. In Paper II, I show that in fact even if only stars were
present, a self-consistent treatment of the stellar SED incorporating
X-rays produced by supernovae favours ${\rm H_{2}}$ formation over
destruction in dense regions.     

\end{itemize}

In this paper, I study the emission mechanisms and observational
signatures of X-ray bright star clusters at high redshift. In
Paper II (Oh 2000a), I address the changes in
reionization topology, reheating and early universe chemistry
mentioned above due to these X-rays. 

In all numerical estimates, I assume a background
cosmology given by the 'concordance' values of Ostriker \& Steinhardt
(1995): $(\Omega_{m},\Omega_{\Lambda},\Omega_{b},h,\sigma_{8
h^{-1}},n)=(0.35,0.65,0.04,0.65,0.87,0.96)$. This corresponds to $\Omega_{b}
h^{2} =0.017$, compared with $\Omega_{b} h^{2}=0.020 \pm 0.002 \, (95
\% {\rm c.l.})$ (Burles, Nollett \& Turner 2000), and $\Omega_{b}
h^{2}=0.0205 \pm 0.0018$ (O'Meara et al 2000) from Big
Bang Nucleosynthesis, and the significantly higher values $0.022 < \Omega_{b}
h^{2} < 0.040 (95 \% {\rm c.l.})$ (Tegmark \& Zaldarriaga 2000)
preferred by recent CMB anisotropy data, such as Boomerang and Maxima.

\section{Emission mechanisms}

\subsection{Star formation at $z > 3$}

What fraction of present day stars formed at high redshift? Estimates
of the comoving star formation rate as a function of redshift (Madau
et al, 1996) should be regarded as lower bounds, particularly at high redshift, due
to the unknown effects of dust extinction, and star formation in faint
systems below the survey detection threshold. Indeed, Lyman break
survey results for $3.6< z <4.5$
(Steidel et al 1999) suggest that after correction for dust
extinction, the comoving star formation rate for $z>1$ is constant,
rather than falling sharply as previously believed. Furthermore, in recent years compelling evidence has emerged that the majority of
stars in ellipticals and bulges formed at high redshift, $z > 3$. This
comes from the tightness of correlations between various global
properties of ellipticals which indicate a very small age dispersion
and thus a high redshift of formation, unless their formation was
synchronized to an implausible degree. The evidence includes the
tightness of the fundamental plane and color magnitude relations for
ellipticals, and the modest shift in zero-points for these relations
with redshift (Renzini 1998 and references therein). Since
spheroids contain $\sim 30\%$ of all stars in the local universe (King
\& Ellis 1985, Schechter \& Dressler 1987),
this would imply that $\sim 30\%$ of all stars have formed at
$z>3$. Since $\sim 20 \%$ of baryons have been processed into stars by
the present day (Fukugita, Hogan \& Peebles 1998), this implies that
$\sim 0.2 \times 0.3 \sim 6 \%$ of baryons have been processed into
stars by $z \sim 3$. Assuming widespread and uniform enrichment and 1
${\rm M_{\odot}}$ of metals per 100 ${\rm M_{\odot}}$ of stars formed,
this translates into an IGM metallicity of $6 \times 10^{-4} \sim 3
\times 10^{-2} Z_{\odot}$. At $z \sim 3$, the metallicity of damped
Ly$\alpha$ systems appears to be $\sim 0.05 Z_{\odot}$ (Pettini et al
1997), in reasonable agreement. The observed metallicity of the Ly$\alpha$
forest at $z \sim 3$, to which most models of reionization have been normalized, is between $10^{-2} Z_{\odot}$ and
$10^{-3} Z_{\odot}$ (Songailia \& Cowie 1996, Songailia 1997), which would imply that only $0.2-2 \%$ of present day stars formed at $z > 3$.  However,
its metallicity may be more representative of low density regions,
rather than the mean cosmological metallicity (Cen \& Ostriker 1999). Note that
normalization of high redshift star formation to Lyman $\alpha$ forest metallicities assume efficient
metal ejection (which underestimates star formation if a significant fraction
of metals are retained) and mixing (which overestimates star formation
if Ly$\alpha$ lines are preferentially observed in overdense regions
which are sites of star formation). In this context, it is worth
mentioning claims that Ly$\alpha$ lines with $10^{13.5} {\rm cm^{-2}} <
{\rm N_{HI}} < 10^{14.5} \, {\rm cm^{-2}}$ reveal lower metallicities
by a factor of 10 than clouds with ${\rm N_{HI}} > 10^{14.5} \, {\rm cm^{-2}}$ (Lu et al
1999).  Ellison et al (2000) find no break in the power law
column density distribution for C IV down to log N(C IV) =11.7, and
Schaye et al (2000) detect O VI down to $\tau_{HI} \sim 10^{-1}$ in
underdense gas, so it appears that metal pollution was fairly
widespread. I regard ${\rm Z} \sim 10^{-3} -2.5 \times 10^{-2} Z_{\odot}$ at $z \sim 3$ as a fairly firm bracket on the range of
possibilities. For inverse Compton radiation, this corresponds to an energy release
per IGM baryon of $\epsilon^{SN} \sim 10
\left( \frac{Z}{10^{-2} Z_{\odot}} \right) \left( \frac{\epsilon}{0.1}
\right)$eV (where $\epsilon$ is the efficiency of conversion of
supernova energy to X-rays), which is comparable to the energy release
in stellar UV radiation for the low escape fractions expected, $\epsilon^{stellar} \sim 10
\left( \frac{Z}{10^{-2} Z_{\odot}} \right) \left( \frac{f_{esc}}{0.01}
\right)$eV, where $f_{esc}$ is the escape fraction of ionizing photons
from the source.  

\subsection{X-ray emission in local starbursts}

Most models of reionization use
population synthesis codes to estimate the spectral energy distribution of
starbursts. However, there are many processes
associated with star formation that generate UV and X-rays, beside
stellar radiation: massive X-ray
binaries, thermal emission from supernova remnants and hot gas in
galactic halos and winds, inverse Compton scattering of soft photons
by relativistic electrons produced in supernovae. Indeed, X-ray
emission appears to be ubiquitous among starbursts (e.g., Rephaeli,
Gruber, \& Persic 1995), and starburst
galaxies may account for a significant portion of the XRB (Bookbinder
et al 1980, Rephaeli et al 1991, Moran, Lehnert \& Helfand 1999). The
X-ray emission from these processes, which hardens the spectrum of starbursts and changes both the topology and chemistry of reionization, has to date been neglected in studies
of the $z > 5 $ universe.  

The X-ray luminosity of starbursts correlates
well with other star formation indicators;  for example, David et al (1992)
find a roughly linear relation between $L_{FIR}$ and $L_{X}$. As a
very rough empirical calibration, the starburst galaxies M82 \& NGC
3256 observed with ROSAT and ASCA (Moran \& Lehnert 1997, Moran, Lehnert \&
Helfand, 1999) follow the relations (after correction for
absorption): $L_{X, 0.2-10 \, {\rm keV}}=8
\times 10^{-4} L_{IR}$, and 
$L_{X, 5 \, {\rm keV}} = 1.2 \times 10^{4} L_{R, 5 \, {\rm GHz}}$ (note that 5 keV flux density is relatively unaffected by photoelectric
absorption or soft thermal emission). The starburst model of
Leitherer and Heckman (1995) yields $L_{bol} \sim L_{{\rm FIR}} \sim 1.5
\times 10^{10} ({\rm SFR}/ 1 \, {\rm M_{\odot} yr^{-1}}) L_{\odot}$; as a
cross-check, the empirical relation for radio emission is (Condon
1992) $L_{R} = 1.4 \times 10^{28} (\nu/{\rm GHz})^{-\alpha} ({\rm SFR}/{\rm M_{\odot}
yr^{-1}}) \, {\rm erg \, s^{-1} \, Hz^{-1}}$, where $\alpha \sim 0.8$. Together I obtain:
\begin{equation}
L_{X}= 5 \times 10^{40} \left( \frac{SFR}{1 \, {\rm M_{\odot} yr^{-1}}}
\right) \ \ {\rm erg \, s^{-1}}
\label{Xray_lum}
\end{equation}
One should not regard this as more than a rough order of magnitude
estimate; a large scatter is expected in this relation. By way of
comparison, Rephaeli et al (1995) obtain from the mean of 51
starbursts observed with Einstein and HEAO, $L_{X,2-30 \, {\rm keV}} \sim 8
\times 10^{-3} L_{IR}$, an order of magnitude greater; and David et al (1992) obtain from a sample of 71 normal and starburst galaxies a ratio lower
by about an order of magnitude, largely due to the inclusion of normal
galaxies (this is consistent with an inverse Compton origin for X-rays, since normal galaxies have much lower radiation field energy densities and would not be expected to show significant inverse Compton emission).   

The typical observed X-ray spectrum is a power-law, $L_{\nu}
\propto \nu^{-0.8}$, consistent with a non-thermal origin. An obscured
AGN is not likely to be the source of these X-rays, as several
observations suggest that the X-ray emission is powered primarily by
massive stars. In NGC 3256, observations by ISO fail to
detect high excitation emission lines (Rigopoulou et al 1996). In M82, the
optical spectrum is HII-like (Kennicutt 1992), discrete
nuclear radio sources are spatially resolved (Muxlow et al 1994), its nuclear X-ray emission
is extended (Bregman et al 1995), and the expected broad H$\alpha$ emission is not
detected (Moran \& Lehnert 1997); the HEX continuum and Fe-K line
emission of NGC 253 as observed by BeppoSAX is extended (Cappi et al
1999). Moran \& Lehnart (1997) and Moran,
Lehnart, \& Helfand (1999) have modelled the X-ray emission of M82 and
NGC 3256, and find inverse-Compton emission to be the most
likely mechanism, rather than an obscured AGN or massive X-ray
binaries. The ratio between the observed radio and X-ray fluxes (note that $L_{X}/L_{syn} \propto
U_{IR}/U_{B}$, where $U_{IR}$ is the energy density of the infra-red
radiation field, and $U_{B}$ is the energy density of the magnetic field) is consistent with an inverse-Compton origin for the
X-rays. Furthermore, the X-ray and radio emission have
the same spectral slope, as is expected if both types of emission are
non-thermal, arising from the same population of electrons. The X-ray luminosity is energetically consistent with a star
formation origin. Assuming a Salpeter IMF and that each supernova explosion yields
$10^{51}$ erg in kinetic energy yields an energy injection rate into
the ISM:
\begin{equation}
\dot{E}_{SN} \sim 3 \times 10^{40} \SFR \efficiency  \, {\rm erg \,
s^{-1}}
\label{IC_lum}
\end{equation}
where $\epsilon$ is the fraction of energy injected into
relativistic electrons; consistency with the observed value,
(\ref{Xray_lum}), implies that $\epsilon \sim 10\%$. Hereafter, I shall use the empirical equation (\ref{Xray_lum}) as a fiducial conversion between X-ray luminosity and star formation rate.

Is an acceleration efficiency of $\epsilon \sim 10\%$ reasonable? The acceleration mechanism for
relativistic electrons is poorly understood. While first-order Fermi acceleration in
shocks is widely accepted as the acceleration mechanism for cosmic
rays (Blandford \& Eichler 1987, Jones \& Ellison 1991), the electron
acceleration is thought to be more problematic, due to the smaller
electron gyroradius (which leads to greater difficulties in bouncing
an electron back and forth across a shock of finite thickness), and the difficulty of initially boosting the
electron to relativistic speeds, where Fermi acceleration can operate
(Levinson 1994). From measurements of cosmic rays energy density it is
inferred that $\sim 10\%$ of the supernova kinetic energy, or $\sim
10^{50}$erg per explosion, is liberated as cosmic rays (Volk, Klein,
\& Wielebinski 1989), but the division between electrons and protons at the
source is not known. Since the measured ratio of cosmic ray protons to
electrons is $\sim 75$ (e.g., Gaisser 1990), it might well be
that relativistic electrons only constitute $\epsilon \sim 10^{-3}$ of the
supernova energy budget. Thus, the reader should be cautioned that $\epsilon \sim 0.1$,
which corresponds roughly equal energy division between protons and
electrons (e/p=1), may be an overly optimistic estimate of the energy
injection into relativistic electrons. Theoretical models of shock
acceleration, in which e/p is a free parameter, often set e/p$\sim 1-5
\%$ for consistency with cosmic-ray experiments (Ellison \& Reynolds
1991, Ellison et al 2000). However, this is somewhat model-dependent:
in models where electrons are injected directly from the thermal pool,
$\sim 5 \%$ of the energy in the
shock must go to non-thermal electrons in order to match gamma-ray
observations (Bykov et al 2000). Furthermore, note that the observed cosmic-ray e/p ratio could equally well be the result
of different transport processes and energy loss mechanisms for
electrons and protons. In particular, cosmic ray electrons are subject
to loss processes which operate on much longer timescales for cosmic
ray protons (inverse Compton, synchrotron losses, etc); the cosmic ray
flux at earth for electrons could arise from a much smaller effective volume
than that for protons. At the source,
the energy division between protons and electrons could range between
1 and 100. Perhaps the most reliable means of inferring the proton/electron energy
division is by direct observations of supernova remnants. In
modelling the observed production of gamma-rays in the supernova
remnants IC 443 and $\gamma$ Cygni observed by the EGRET instrument on
the Compton Gamma Ray Observatory, Gaisser, Protheroe \& Stanev (1998)
find that a proton to electron ratio of 3--5 gives the best fit to the
observed spectra, implying $\epsilon \sim 0.02-0.03$. Similarly, in
modelling the $\gamma$-ray flux from 2EG J1857+0118 associated with
supernova remnant W44, de Jager \& Mastichiadis (1997) find $\epsilon
\sim 0.09$. They speculate that electron injection by the pulsar may
be responsible for the increased electron energy content. Given the
large uncertainties, henceforth I shall simply use the empirical
relation (\ref{Xray_lum}).

Thus, barring
non-standard IMFs, type II detonation energies or alternate sources of
relativistic electrons, the empirical relation (\ref{Xray_lum}) implies an
acceleration efficiency $\epsilon \sim 0.1$ which is 
plausible but certainly lies at the upper limit of theoretical
expectations. Another possibility is that the X-ray emission cannot be
wholly attributed to inverse Compton emission alone (this assumption rests on the arguments of Moran \&
Lehnart (1997) and Moran, Lehnart, \& Helfand (1999) with regards to the
slopes and relative intensities of the observed non-thermal radio and X-ray
emission). The X-ray emission may instead be due to X-ray binaries,
thermal emission from supernova remnants or starburst driven
superwinds (e.g., see Natarajan \& Almaini 2000). It should be noted that only soft X-rays are relevant
for reionization, since the universe is optically thin to photons with
energies $E > E_{thick}=1.5 \redshift^{0.5} x_{HI}^{1/3} \, {\rm keV}$ (where
$x_{HI}$ is the mean neutral fraction of the IGM). Since equation
(\ref{Xray_lum}) is calibrated with the soft bands observed by ROSAT,
this implies that even if $\epsilon \ll 0.1$ and the observed X-rays are not
predominantly due to inverse Compton emission,
the importance of X-rays for reionization (in particular, for changing
the topology, for increased reheating, and increased $H_{2}$
production) may still hold. However, observational predictions
which focus specifically on the inverse Compton mechanism (e.g., the
gamma-ray background (section (\ref{xray_obs})), and detecting synchrotron emission with
the SKA (section (\ref{radio_obs})) will no longer be valid. Since the
acceleration efficiency is the most uncertain parameter in this paper,
wherever relevant I insert the scaling factor $\efficiency$ into
numerical estimates.   

How does the X-ray luminosity compare with stellar UV ionizing radiation? Assuming a Salpeter IMF with solar metallicity, the
Bruzual \& Charlot (1999) population synthesis code yields an energy
output of $L_{ion}= 3.2 \times 10^{42} ({\rm SFR/1
{\rm M_{\odot} \, yr^{-1}})\, {\rm erg \, s^{-1}}} $ in ionizing
photons, which translates into ${\rm \dot{N}_{ion}} = 10^{53} ({\rm SFR}/1
{\rm M_{\odot} \, yr^{-1}}) \, {\rm photons \, s^{-1}}$. However, note that
most of these ionizing photons are absorbed locally with the ISM of
the star cluster; the escape fraction of ionizing photons into the IGM
is expected to be small. Leitherer et al (1995) have observed four starburst galaxies with the Hopkins
Ultraviolet Telescope (HUT). Their analyis suggests an escape fraction of only
$3\%$, based on a comparison between the observed Lyman continuum flux
and theoretical spectral energy distributions. For our own Galaxy,
Dove, Shull \& Ferrara (2000) find an
escape fraction for ionizing photons of $6\%$ and $3\%$ (for coeval and
Gaussian star formation histories respectively) from OB associations
in the Milky Way disk. Bland-Hawthorn \& Maloney (1999) find an escape fraction of $6\%$ is necessary for consistency with the observed H$\alpha$ emission
from the Magellanic stream and high velocity clouds. On the other
hand, a recent composite spectrum of 29 Lyman break galaxies (LBGs) with
redshifts $\langle z \rangle = 3.40 \pm 0.09$ shows significant
detection of Lyman continuum flux (Steidel, Pettini \& Adelberger
2000); for typical stellar synthesis models, the observed flux ratio L(1500)/L(900)=$4.6 \pm 1.0$ implies
little or no photoelectric absorption. The fraction of 900 ${\rm \AA}$
photons which escape, $f_{esc} \sim 15 -20 \%$, is modulated almost
entirely by dust absorption. Nonetheless, the authors themselves stress
this result should be treated as preliminary; the result could be due
to a large number of uncertainties or selection effects, among them
the fact that these galaxies were selected from the bluest quartile of
LBGs. On the
theoretical side, radiative transfer calculations by Woods \& Loeb (1999) find that the escape fraction
at $z \sim 10$ is $< 1\%$ for stars; calculations by Ricotti \& Shull
(1999) find that the escape fraction decreases strongly with
increasing redshift
and halo mass; for a $10^{9} \, M_{\odot}$ halo at $z=9$, the escape
fraction is $\sim 10^{-3}$ (note that in the Ricotti \& Shull (1999)
models, the escape fraction rises towards low masses, and can be
considerable for the halos with $M < 10^{7} M_{\odot}$. Since such
halos have $T_{vir} < 10^{4}$K, their contribution to reionization
depends on whether $H_{2}$ formation and cooling can take place
despite photodissociative processes (Haiman, Rees \& Loeb 1997, Cicardi, Ferrara
\& Abel 1998, Ricotti, Gnedin \& Shull 2000)). For low escape fractions, the energy
release in UV photons is roughly comparable to that in inverse-Compton X-rays :
\begin{equation}
L_{UV} = 3 \times 10^{40} \left( \frac{f_{esc}}{0.01} \right) \left
( \frac{\rm SFR}{1 {\rm M_{\odot} yr^{-1}}} \right) \, {\rm erg \, s^{-1}}
\label{lum_stars}
\end{equation} 
Note that to first order the ratio of stellar UV to inverse Compton
X-rays is not sensitive to uncertainties in the IMF, as the same massive stars with $M > 20
M_{\odot}$ that dominate the Lyman continuum of a stellar population
also explode as supernovae (however, see section (\ref{metal_free})
for some caveats).  

Will X-ray emission still be efficient at high redshift?
The following changes are expected to take place at high
redshift: (i) The ISM is initially free of dust and metals, although
rapid enrichment could occur on fairly short timescales ($t \sim
10^{6}-10^{7}$yr). (ii) The average ISM density
is significantly higher, $n_{halo} \sim n_{0} 18 \pi^{2} (1+z)^{3} = 0.02
(\frac{1+z}{10})^{3} {\rm cm^{-3}}$ in a
halo and $n_{disc} \sim \lambda^{-3} n_{halo} \sim 160 
(\frac{1+z}{10})^{3} {\rm cm^{-3}} (\lambda/0.05)^{-3}$ in a disk (where $\lambda$ is the
spin parameter). A supernova remnant at high $z$ expands into a denser
ISM: since it spends a shorter time in the Taylor-Sedov phase, most of
its energy is radiated at a smaller radius, where the effective
temperature is higher. This implies a harder spectrum for thermal emission. In addition, the density and
temperature of gas in star forming regions is determined by the
properties of ${\rm H}_{2}$ cooling (which saturates at $n \sim 10^{4} \,
{\rm cm^{-3}}$ and $T \sim 300$K), rather than metal cooling as in the
local universe. (iii) Potential wells are significantly
shallower, so pressurised regions (hot gas, strong magnetic fields)
cannot be efficiently confined. (iv) The CMB energy density $U_{CMB}
\propto (1+z)^{4}$, so inverse Compton radiation becomes particularly
efficient at high redshift. Because of (iii) and (iv), relativistic electrons cool predominantly by inverse Compton scattering rather than synchrotron emission. Since inverse Compton emission is likely to be the most promising mechanism for X-ray emission, I shall consider it at length.  

\subsection{Inverse Compton emission}

The energy loss rate for a relativistic electron with Lorentz factor
 $\gamma$ is given by:
\begin{equation}
\dot{{\rm E}_{IC}}= \frac{4}{3} \sigma_{T} c \gamma^{2} U_{rad} = 1.12
\times 10^{-16} \redshift^{4} \left( \frac {\gamma}{10^{3}}
\right)^{2} \ {\rm erg \, s^{-1}}
\label{IC_rate}
\end{equation}
for $U_{rad}= U_{\rm CMB}$. The loss rate by synchrotron radiation is given by substiting
 $U_{B}$ for $U_{rad}$, $\dot{{\rm E}_{synch}}= \frac{4}{3} \sigma_{T}
 c \gamma^{2} U_{B}$. In the local universe, galaxies with relatively quiescent star formation
 emit most of their electron energy in synchrotron
 radiation. Starbursts in the local universe can radiate efficiently
 in IC, as the energy density in the local radiation field is sufficiently high (typically, $U_{r} \sim 10^{-8} \, {\rm erg \, cm^{-3}}$ as opposed to $U_{r} \sim 10^{-12} \, {\rm erg \, cm^{-3}}$
in our Galaxy). Seed photons are provided by IR emission from dust
 grains. By contrast, at high redshift, {\it all} star forming regions will emit in inverse Compton radiation, as the CMB
 provides a universal soft photon bath of high energy density,
 $U_{CMB} = 4 \times 10^{-9} (\frac{1+z}{10})^{4}\, {\rm erg \,
 cm^{-3}}$. An electron with Lorentz factor $\gamma$ will boost a
CMB photon of frequency $\nu_{o}$ to a frequency $\nu= \gamma^{2}
\nu_{o}$. Thus, the rest frame frequency of a CMB photon (at the peak
of the blackbody spectrum) which undergoes inverse Compton scattering is:
\begin{equation}
{\rm E}_{IC}=600 \left( \frac{\gamma}{300} \right)^{2} \left( \frac{1+z}
{10} \right) \, {\rm eV} 
\end{equation}
Note that the observed frequency is independent of source redshift,
since the higher initial frequency and redshifting effects cancel out.

Below, I examine in detail the mechanisms by which
 relativistic electrons lose energy, to see if indeed inverse Compton
 radiation will predominate.  

\subsubsection{Energy loss processes for relativistic electrons}

Once relativistic electrons are produced, they can cool via a variety of
mechanisms. Let us examine them in turn (for more details see Pacholczyk 1970, Daly 1992). 

{\bf Synchrotron radiation} The relative emission rate in inverse
Compton and synchrotron emission is given simply by the relative
energy densities in the radiation and magnetic fields,
$\dot{E}_{IC}/\dot{E}_{syn}= U_{rad}/U_{m}$. This yields:
\begin{equation}
\frac{\dot{E}_{IC}}{\dot{E}_{syn}}= 1.1 \times 10^{3} \left( \frac{B}
{10 \mu G} \right)^{-2} \left( \frac{1+z}{10} \right)^{4}  
\end{equation}
Note that in energy loss terms the CMB may be characterized as having
an effective magnetic field strength $B_{CMB,eff}=3.24 \times 10^{2}
\left( \frac{1+z} {10} \right)^{2} \ {\rm \mu G}$. Could magnetic
fields in proto-galaxies possibly reach these high values? A reasonal
assumption is that $P_{B} \sim P_{rel} < P_{gas}$, where $P_{B}$ is the
magnetic field pressure, $P_{rel}$ is the pressure in relativistic
particles, and $P_{gas}$ is the thermal gas pressure. Local observations of
synchrotron and inverse Compton emission from radio galaxies are
consistent with equipartition $P_{B} \sim P_{rel}$ (Kaneda et al
1995). I have assumed that the energy injection into
relativistic particles is $\sim 10 \%$ of the total kinetic energy of
a supernova, so $P_{rel} < P_{gas}$ should be a strict upper bound. Thus, $P_{B} < P_{gas}$ gives
the upper bound:
\begin{equation}
B < 6 \left( \frac{n}{1 {\rm cm^{-3}}} \right)^{1/2} \left( \frac{T}{10^{4}
K} \right) \, \mu G
\end{equation} 
where $n$ is the baryon number density (note that gas with temperatures $> 10^{4-5}$K will escape
from the shallow potential wells of the first proto-galaxies). If the
magnetic field exceeds the above value, the over-pressurised lobe
will expand on the dynamical time scale until the magnetic pressure drops. Thus, for $z>5$, it
seems likely that synchrotron energy losses will be unimportant.

{\bf Ionization \& Cherenkov losses} Interactions with the
non-relativistic gas will result in energy losses via ionization and
Cherenkov emission of plasma waves at a rate independent of the
Lorentz factor, $\dot{E}_{ion} \approx 9 \times 10^{-19} n \, {\rm
erg \, s^{-1}}$, which implies that the relative energy loss rate is:
\begin{equation}
\frac{\dot{E}_{IC}}{\dot{E}_{ion}} = 120 \left( \frac{n}{1 {\rm cm^{-3}}}
\right)^{-1} \left( \frac{1+z}{10} \right)^{4} \left( \frac
{\gamma}{10^{3}} \right)^{2}
\end{equation}
Thus, for 
\begin{equation}
\gamma > \gamma_{break} \approx 100  \left( \frac{n}{1 {\rm cm}^{-3}}
\right)^{1/2} (\frac{1+z}{10})^{-2}
\label{gamma_break}
\end{equation}
inverse Compton losses are more important than ionization losses. The cutoff Lorenz factor at the lower end $\gamma_{co}$, is of interest since
the lower end accounts for most of the electrons, both in terms of
number and energy: $N_{electrons} (> \gamma) 
\propto \gamma^{-2 \alpha} = \gamma_{co}^{-1.6} \ $; $E_{electrons} (> \gamma)
\propto \gamma^{-2 \alpha+1} = \gamma_{co}^{-0.6} \ (\alpha=0.8)  $. Note that
the electrons that cool via ionization losses do not drop out
completely, but merely form a flattened distribution with $\gamma_{f}
\approx \gamma_{i} -350 \, n ({\rm t/10^{7} yr})$. These electrons with
lower Lorentz factors could scatter CMB photons to optical and UV
frequencies.  

{\bf Free-free radiation} Free-free radiation results in an energy
loss rate $\dot{E}_{free-free} = 6 \times 10^{-22} n \gamma \,
{\rm erg \, s^{-1}}$. Thus, free-free radiation only dominates over
ionization and Cherenkov radation for $\gamma > 1500$. However, in
this regime inverse Compton losses dominate, since
\begin{equation}
\frac{\dot{E}_{IC}}{\dot{E}_{ff}}= 190 \left( \frac{n}{1 cm^{-3}}
\right)^{-1} \left( \frac{1+z}{10} \right)^{4} \left( \frac
{\gamma}{10^{3}} \right)
\end{equation}
Thus, free-free emission is never important in cooling relativistic
electrons at high redshift.

In summary, the inverse Compton emission is the dominant energy loss
mechanism between a lower and upper energy cutoff. From equation
(\ref{gamma_break}), the
lower frequency break is determined by the competition between the inverse
Compton loss rate and ionization and atomic cooling losses, below
\begin{equation}
E_{lower} = \gamma_{break}^{2} h \nu_{{\rm CMB}}= 70 \density  \left(
\frac{1+z}{10} \right)^{-3} \, {\rm eV},
\end{equation} 
where I assume the seed photon $\nu_{{\rm CMB}}=1.6 \times 10^{12} \redshift$GHz lies
at the peak of the CMB blackbody spectrum. The competition between
inverse Compton cooling losses and the rate of energy injection by
Fermi acceleration determines the upper energy cutoff. The timescale
for losses by inverse Compton radiation is:
\begin{equation}
t_{life}= \frac{E}{\dot{E}} = 7.9 \times 10^{5} \left( \frac{1+z}{10}
\right)^{-4} \left( \frac{\bar{\gamma}}{300} \right)^{-1}  \ {\rm
years}
\label{lifetime}
\end{equation} 
Equating
the Fermi acceleration timescale $t_{acc} \sim r_{L}c/v_{sh}^{2}= 1.3
\times 10^{-3} \left( \frac{\gamma}{300} \right) \left( \frac{B}{10
\mu G} \right)^{-1} \left( \frac{v_{sh}}{2000 {\rm km s^{-1}}}
\right)^{-2}$yr (where $r_{L}$ is the Larmour gyroradius, and $v_{sh}$
is the typical shock velocity) to $t_{life}$ (equation
(\ref{lifetime}), I obtain for the maximum Lorentz factor
$\gamma_{max}= 7.4 \times 10^{6} \Bfield^{1/2} \left(
\frac{v_{sh}}{2000 \, {\rm km \, s^{-1}}} \right) \redshift^{-2}$
which corresponds to an upper energy cutoff:
\begin{equation}
E_{upper} \sim 360 \Bfield \left(
\frac{v_{sh}}{2000 {\rm km \, s^{-1}}} \right)^{2} \redshift^{-3} \, {\rm
GeV}.
\label{Emax_eqn}
\end{equation} 
In section (\ref{IC_spectrum_section}), I derive the
form of the spectrum. For the power law spectrum obtained, $L_{\nu} \propto
\nu^{-1}$, the specific luminosity depends only logarithmically on the
energy cutoffs: $L_{\nu} = \frac{L_{tot}}{{\rm log
(\nu_{upper}/\nu_{lower}})} \nu^{-1}$. 
                                                                                   
\subsubsection{Inverse Compton spectrum}
\label{IC_spectrum_section}

The Fermi shock acceleration mechanism for cosmic rays involves a steady growth of particle
energy as a particle scatters back and forth across the shock front. This
naturally produces a power law electron energy spectrum with energy
spectrum $dn/d\gamma \propto \gamma^{-p}$, where $p=(\chi+2)/(\chi-1)$
and $\chi$ is the compression ratio for the shock (e.g., Jones \&
Ellison 1991). The final emission spectrum $L_{\nu} \propto
\nu^{-\alpha}$ depends on the electron energy spectrum; the
synchrotron or inverse Compton 
spectral index is $\alpha=(p-1)/2$. Thus, the spectral
index is determined by the shock structure rather than the details of
the scattering process. For a strong
adiabatic shock $\chi =4$ and $dn/d\gamma \propto \gamma^{-2}$; for an
isothermal shock, $\chi \gg 1$ and $dn/d\gamma \propto
\gamma^{-1}$. The assumption of adiabaticity is most appropriate for
the Taylor-Sedov phase, when most of the electrons are
accelerated. Galactic SNR show a mean radio spectral index
$\alpha= 0.5 \pm 0.15$ (Droge et al 1987), which agrees with
$dn/d\gamma \propto \gamma^{-2}$. I shall use this as the canonical
electron injection spectrum in this paper. The steepening of the
diffuse synchrotron emission to $\alpha \sim 0.8$ is likely to be
due to energy losses as the electrons age. I examine the emission
spectrum in detail below. I assume that there is one supernova for
every 100 $M_{\odot}$ of stars formed, that each supernova liberates
$\sim 10^{51}$erg in kinetic energy, and $\sim 1 M_{\odot}$ of
metals.

The equation for the evolution of the electron population is given by
(Ginzburg \& Syrovatskii 1964, Sarazin 1999):
\begin{equation}
\frac{ \partial N(\gamma)}{\partial t} = \frac{\partial}{\partial
\gamma} [b(\gamma)N(\gamma)] + Q(\gamma)
\label{electron_pop_eqn}
\end{equation}
where $N(\gamma)$ is the number of electrons in the range $\gamma$ to
$\gamma + d\gamma$, the rate of production of new relavistic electrons
is given by $Q(\gamma)$, and the rate of energy loss of an individual
particle is given by $b(\gamma) \equiv d\gamma/dt$. From equation
(\ref{lifetime}), we see that the electron lifetime is shorter than
both the Hubble time for the redshifts under consideration $z < 30$ and typical
timescales for starbursts ($\sim 10^{7}$ years). We can thus assume
that the electron population quickly reaches a steady state where the
energy losses due to inverse Compton scattering (for high $\gamma$)
and Coulomb collisions (for low $\gamma$) balance the injection rate
due to supernova explosions, and set the time derivative to zero. Since injected electrons survive for less than a Hubble time, I ignore evolution in loss rate due to the evolution of the CMB energy
density. Assuming that each supernova injects a population of
electrons with $Q(\gamma) = Q_{o} \gamma^{-p}$, where $p \sim 2$, then the steady state solution to equation
(\ref{electron_pop_eqn}) is given by:
\begin{eqnarray}
N(\gamma) &=& 4 \times 10^{61} \redshift^{-4}  \SFR \gamma^{-(p+1)} \ \ \ ; \gamma >
\gamma_{break}
\label{electron_pop_soln}  \\ 
N(\gamma) &=& 4 \times 10^{57} \density^{-1} \SFR \gamma^{-(p-1)} \ \ \ ; \gamma<
\gamma_{break} 
\nonumber 
\end{eqnarray}   
where $\gamma_{break}$ (equation (\ref{gamma_break})) is the transition energy between the regimes
where ionization and inverse Compton losses dominate. Thus, the electron
population flattens by one power at low energies and steepens by one
power at high energies, as compared with the distribution function for
the injected population. Since 
$\alpha=-(p-1)/2$, the
emitted spectrum flattens by a half power and steepens by a half power
at low and high energies respectively. In particular, for $p=2$, as is
appropriate for adiabatic shocks, $L_{\nu} \propto {\rm const}$ at low
energies and $L_{\nu} \propto \nu^{-1}$ at high energies. 

\begin{figure}
\epsscale{1.00}
\plotone{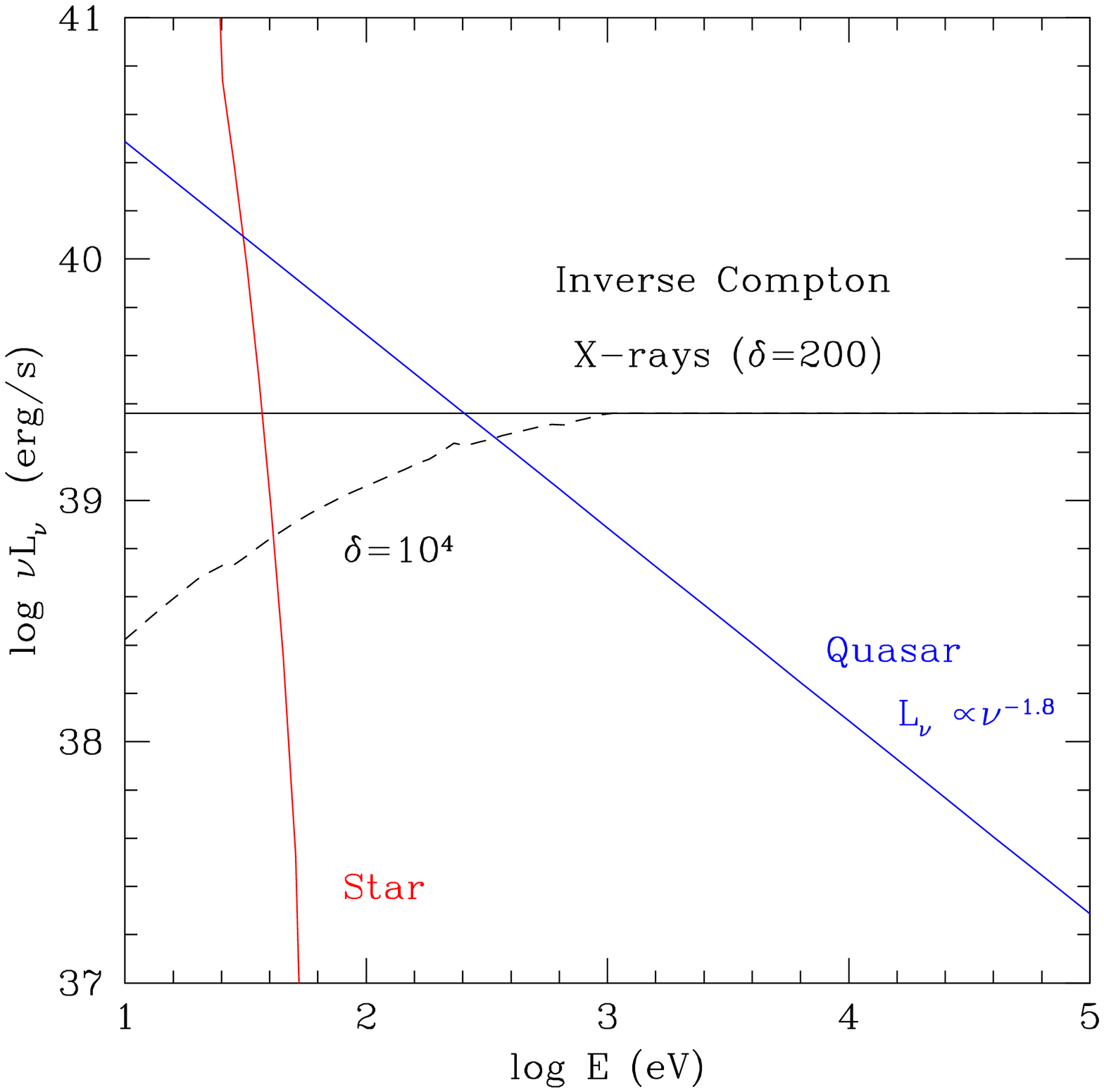}
\caption{Comparison of stellar, QSO (with $L_{\nu} \propto
\nu^{-1.8}$) and inverse Compton spectral energy distributions, all of
which are normalised to have the same luminosity above 1 Ry. The
inverse Compton SED is significantly harder than all others.  At high densities
(overdensities $\delta \sim 10^{4}$ at $z \sim 9$), cooling of relativistic
electrons by ionization losses becomes significant, causing the
downturn of the spectrum at low energies.
\label{spectrum}}
\end{figure}

The spectrum of inverse Compton radiation is given by (Sarazin 1999): 
\begin{equation}
L_{\nu} = 12 \pi \sigma_{T} \int_{1}^{\infty} N(\gamma) d \gamma
 \int_{0}^{1} J(\frac{\nu}{4 \gamma^{2} x}) F(x) dx 
\label{IC_spectrum}
\end{equation}
where $N(\gamma)$ is the electron energy spectrum, $J(\nu)=
B_{\nu}(T_{CMB,o}(1+z))$ is the CMB blackbody spectrum, and 
$F(x)=1+x+2x {\rm ln}(x) -2 x^{2}$. This differs only by a
normalization correction from assuming that all photons reside at the peak of the
blackbody spectrum, $\nu = 1.6 \times 10^{11} (1+z)$Hz. This is
because the seed photon spectrum is narrow compared to the electron energy
distribution. In figure (\ref{spectrum}),
I show example spectra computed with equations (\ref{electron_pop_soln})
and (\ref{IC_spectrum}). Note that the efficiency of Coulomb cooling
depends on the assumed electron density; I assume that $n_{e} \sim
\delta n_{o} (1+z)^{3}$, where I assume the overdensity to be either
$\delta \sim 200$ (to mimic the overdensity at the virial radius) or
$\delta \sim 10^{4}$ (to mimic the overdensity for collapsed gas in
dense star forming regions). At $z\sim 10$, this assumption
corresponds to $n_{e} \sim 4 \times 10^{-2} \, {\rm cm^{-3}}$ and
$n_{e} \sim 2 \ {\rm cm^{-3}}$ respectively. Also plotted for reference is the
spectrum of the same starburst with a Salpeter IMF, assuming an escape fraction of $\sim 1 \%$,
and the spectrum of a mini-quasar with spectrum $L_{\nu} \propto
\nu^{-1.8}$, normalized to have the same energy release above the Lyman limit as the
starburst. Note that the inverse Compton case has the hardest spectrum
of all. Since the spectrum is so hard, photoelectric absorption by the
host galaxy does not
significantly attenuate the ionizing flux.  Most UV photons produced will not escape the host galaxy, whereas X-ray photons with 
$E > 270 \left( \frac {N_{HI}}{10^{21} cm^{-2}} \right) \,
{\rm eV} $ where $N_{HI}$ is the column density in the host galaxy, will escape
unimpeded. Since $\nu L_{\nu} \sim {\rm const}$, most of the
energy in ionizing photons escape. Note, incidentally, that the escape fraction for UV photons produced by inverse Compton may be significantly higher
than photons of the same frequencies produced by stars, as
relativistic electrons can disperse from the star forming region,
where gas densities are highest and most photon captures take place.    

\subsection{Zero-metallicity star formation}
\label{metal_free}

The stellar IMF at high redshift under conditions of low or zero
metallicity is unknown. Up to now, I have assumed a Salpeter IMF, in
which one supernova expodes for every $\sim 100 \, M_{\odot}$ of stars
formed, and each supernova deposits on average
$\sim 10^{50} \efficiency$erg in relativistic electrons and $\sim 1 \,
M_{\odot}$ of metals. Since the estimated amount of star formation at
high redshift is calibrated to the observed IGM metallicity at $ z
\sim 3$, it is worth asking whether the energy injected into X-rays per
solar mass of metals produced, $\epsilon_{Z} \sim 10^{50} \, {\rm erg \,
M_{\odot}^{-1}}$, could change significantly at high redshift. I have
argued that variations in the IMF will not effect significant changes
in $\epsilon_{Z}$ or the ratio of energy emitted in stellar UV
ionizing photons to inverse-Compton X-rays, since the same massive OB stars which produce
UV ionizing photons also explode as supernova, producing both relativistic
electrons and metals.

However, zero-metallicity star formation may result in very different
stellar populations from that seen at low redshift. The lack of
efficient cooling mechanisms could result in extremely top heavy
stellar IMFs (Larson 1998, Larson 1999) and in particular the
production of ``Very Massive Objects'' (VMOs) in the range
$10^{2}-10^{5} \, M_{\odot}$ (Carr, Bond \& Arnett 1984). Such stars
have high effective temperatures and produce a much harder spectrum
than ordinary stars (Tumlinson \& Shull, 2000, Bromm et al, 2000). Moreover, the
distribution of stellar endpoints is very different from a normal IMF
(Heger, Woosley \& Waters 2000). Stars with masses between 10 and 35
$M_{\odot}$ explode as type II supernovae. While it is not known
whether metal free stars with masses between 35 and 100
$M_{\odot}$ will explode--they might collapse to form black holes--stars with masses $\sim 100-250 M_{\odot}$ are
disrupted by the pair production instability, once again producing an
energetic supernova event and dispersing metals. Stars more massive
than $250 M_{\odot}$ should collapse completely to black holes,
without ejecting any metals (unless they eject their envelopes
during hydrogen shell burning, in which case it is possible for them
to explode). The latter provides an obvious mechanism for seeding
supermassive black holes to form AGN. A number of studies of
zero-metallicity star formation suggest that the Jeans/Bonner-Ebert
mass and hence the lower mass cutoff for star formation is very high,
$M_{*} > 10^{2}-10^{3} \, M_{\odot}$ (Abel, Bryan \& Norman 1999,
Padoan, Nordlund \& Jones 1997).   

There are three main observations to make: (i) VMOs with initial
stellar masses $100 M_{\odot} < M < 250
M_{\odot}$ which are disrupted by the pair-production stability  show $\epsilon_{Z}(PP) \approx
\epsilon_{Z}({\rm type \ II})$, which imply that even if these objects are
abundant in the early universe, our estimates of the level of inverse
Compton X-ray emission do not change strongly. In particular, $E_{\rm explosion}= 6.3 \times 10^{52} \left( \frac{M}{10^{2} M_{\odot}}
\right)^{2.8} \frac{{\rm min}[0.13,(1-\phi_{L})^{2.8}]}{0.13} \, {\rm
erg}$ while the yield in elements heavier than helium is $Z_{\rm ej} =
{\rm min}\left[ (1- \phi_{L}),0.5 \right]$, where $\phi_{L}$ is the fraction
of the initial mass lost during hydrogen burning (Carr, Bond \&
Arnett, 1984). Thus, $\epsilon_{Z}(PP)= 1.3 \times 10^{50} \left( \frac{M}{10^{2} M_{\odot}}
\right)^{1.8} \frac{{\rm min}[0.25,(1-\phi_{L})^{1.8}]}{0.25} \, {\rm
erg \ M_{\odot}^{-1}} \approx \epsilon_{Z}({\rm type \ II})$. (ii) Stars which directly collapse to
form black holes represent an additional, unaccounted source of
ionizing photons, since they are not included in the metal pollution
budget (of course, their most important contribution could lie in
seeding AGN formation). (iii) Zero metallicity is a singularity: true
zero-metallicity stellar populations differ
greatly from low metallicity ones. Even the
introduction of trace amounts of metals $Z \sim 10^{-4} Z_{\odot}$
introduces important changes in stellar structure and evolution
(Heger, Woosley \& Waters 2000). Furthermore, trace amounts of metals drastically reduces the
abundance of VMOs by allowing efficient cooling past the 300K barrier
imposed by ${\rm H}_{2}$ cooling, reducing the Jeans mass and thus the
minimum mass for star formation. Prompt initial enrichment is quite
plausible: the lifetime of massive
stars before they explode to pollute the IGM with metals is $\sim 10^{6}$ years, which is a short fraction of the Hubble
time $t_{H} \sim 8 \times 10^{8} (\frac{1+z}{10})^{-3/2}$yrs even at high
redshift. Although the degree of
metal mixing is uncertain (e.g. Gnedin \& Ostriker 1997, Nath \& Trentham 1997,
Ferrara, Pettini \& Shchekinov 2000), note that first star forming regions are very highly biased, and subsequent
generations of the halo hierarchy collapse in proto-clusters and
filaments very close to the first star clusters. Thus, stars of finite metallicity could quickly predominate, even if the mean
metallicity of the universe is close to primordial (Cen \& Ostriker 1999). It is therefore possible and even likely
that true zero metallicity star formation was confined to a very small
fraction of the stars formed at high redshift, and thus negligible in
terms of the energy budget for reionization.    

\section{Observational signatures}

In this section, to estimate number counts I use the Press-Schechter based
high redshift star formation models of Haiman \& Loeb (1997), which are normalised to the
observed metallicity of the z=3 IGM, ${\rm Z = 10^{-3}-10^{-2}
Z_{\odot}}$. In these models, in every halo capable of atomic cooling
(i.e., with a virial temperature $T_{vir} > 10^{4}$K), a fixed fraction of the gas
$f_{star}=1.7, 17 \%$ (for ${\rm Z}(z=3)= 10^{-3}, 10^{-2} {\rm Z_{\odot}}$
respectively) fragments in a starburst lasting $t_{o} \sim
10^{7}$yrs. The correspondence between halo mass and star formation
rate is thus ${\rm SFR} \approx 2 \left( {\rm \frac{M_{halo}} {10^{9}
\, M_{\odot}} } \right) \left( \frac{f_{star}}{0.17} \right) \
{\rm M_{\odot} \, yr^{-1}}$, and each halo is only visible for some
fraction of the time $\frac{t_{o}}{t_{H}(z)}$. 

\subsection{HeII recombination lines}

One possible signature of the hard spectrum produced by inverse
Compton X-rays would be HeII recombination lines from the host
galaxy. Indeed, such lines may well be detectable from the first
luminous objects with sufficiently hard spectra, such as mini-quasars or metal-free
stars (Oh, Haiman \& Rees 2000). The different sources may perhaps be
distinguished on the basis of line widths and line ratios (Tumlinson,
Giroux \& Shull, 2000). However, inverse Compton emission produces too
few HeII ionizing photons for such recombination lines to be
detectable with NGST at high redshift. For a Salpeter IMF, $\dot{N}_{ion,HI}
= 10^{53} \SFR \ {\rm photons \, s^{-1}}$ from stellar UV
radiation. On the other hand, since secondary ionizations of HeII are negligible (Shull \& van Steenberg
1985), the production rate of HeII ionizing photons from non-thermal
emission is $\dot{N}_{ion, HeII}= \int_{4 \nu_{L}}^{\nu_{thin}} \frac{L_{\nu}}{h \nu} \approx
2.5 \times 10^{49} \left( \frac{SFR}{ 1 \, M_{\odot} \, yr^{-1}} \right) \ {\rm photon
\, s^{-1}}$, where $\nu_{thin}$ is the frequency at which the halo
becomes optically thin. Thus, $Q \equiv \dot{N}^{\rm HeII}_{\rm ion}/\dot{N}^{\rm
HI}_{\rm ion} \approx 2.5 \times 10^{-4}$, as compared with $Q \approx
0.05$ for stellar emission from metal-free stellar population, and $Q
\approx 4^{-\alpha} = 0.08-0.25$ for a QSO, where $L_{\nu} \propto
\nu^{-\alpha}$ and $\alpha = 1-1.8$, implying that the relative flux
in HeII and H$\alpha,\beta$ recombination lines is much smaller for
inverse Compton emission than metal-free stars or AGN. The luminosity in a helium
recombination line $i$ may be estimated as $L_{\rm i} = Q f_{\rm i} L_{\rm
H\alpha}$, where $L_{\rm H\alpha}$ is the Balmer $\alpha$ line luminosity,
$f_{\rm i}\equiv \frac{j_{i}}{j_{H \alpha}} \left
( \frac{\alpha_{B}(HI) n_{HII}}{\alpha_{B}(HeII) n_{HeIII}} \right)$,
$\epsilon$ is the fraction of SN energy which emerges as IC emission, 
and $j_{i},j_{H \alpha}$ may be obtained from Seaton (1978). The observed flux is $J_{i}= \frac{L_{\rm i}}{4 \pi d_{\rm L}^{2}} 
\frac{1}{\delta \nu} = 0.04 \left( \frac{q_{\rm i}}{0.5} \right)  \left(\frac{1+z}{10} \right)^{-1} \left(\frac{\epsilon}{0.1} \right) \left(
\frac{R} {1000} \right) \left( \frac{{\rm SFR}}{1 {\rm M_{\odot} yr^{-1}}}
\right) \, {\rm nJy} $
where ${\rm R}$ is the spectral resolution, and $q_{\rm i} \equiv
f_{\rm i} \nu_{\rm H\alpha}/ \nu_{\rm i}$, where $f_{\lambda
4686}=0.74, f_{\lambda 1640}=4.7, f_{\lambda 3203}=0.30$. This is undetectable by
NGST, which in $10^{5}$s integration time requires for a 10 $\sigma$
detection a flux $F(10\sigma) \sim 30$nJy at these frequencies (observed wavelengths
$1 < \lambda < 5.5 \mu$m) and spectral
resolution $R \sim 1000$ (see Oh, Haiman \& Rees 2000 for details).   

It is worth mentioning that a hard source may produce relatively little
recombination line flux and yet play an important role in reionizing
the universe. The recombination line flux from a source is $\propto
(1-f_{esc})$, whereas the ionizing flux escaping into the IGM is $\propto
f_{esc}$ (where $f_{esc}$ is significantly higher for hard
sources: all photons with $E > E_{halo, thin} = 270
\left(\frac{N_{HI}}{10^{21} {\rm cm}^{-2}} \right)$eV can
escape freely from the host). Secondary ionizations are unimportant
within a host halo: much of the gas in fully ionized, in which case
the energetic
photoelectron deposits its energy as heat. In any case, the halo is
optically thin to the most energetic photons $E > E_{halo,thin}$ which are
important for secondary ionizations. Thus, in recombination line flux what matters is the
total {\it number} of ionizing photons produced, whereas for the
reionization of the IGM what matters is the total output {\it energy} (since
in a largely neutral medium with $x_{e} < 0.1$ the total number of ionizations for a
photon of energy $E_{photon}$, including
secondary ionizations, is $N_{ion} \sim E_{photon}/37 {\rm eV}$ (Shull \& van
Steenberg 1985). Note, however, that as the medium becomes
more ionised an increasing fraction of the energy is deposited as heat, and an
additional source of soft photons is needed for reionization to
proceed (Oh 2000a)). Thus, for instance, for $f_{esc} \sim 1 \%$, inverse
Compton radiation makes a negligible contribution to the H$\alpha$
luminosity, $\frac{L_{H \alpha}^{IC}}{L_{H \alpha}^{stellar}} \approx
\frac{\dot{N}_{ion}^{IC}}{\dot{N}_{ion}^{stellar}} = 2.5 \times
10^{-4}$, but the ionizing IC and stellar radiation escaping from the
host are energetically comparable (from equations
(\ref{Xray_lum}) and (\ref{lum_stars}), $L_{X}^{IC} \approx
L_{UV}^{stellar}$), and thus they produce roughly equal number of
ionizations in the IGM.

\subsection{X-rays and gamma-rays}
\label{xray_obs}

Unfortunately, direct detection of inverse Compton X-rays from
high-redshift star clusters is unlikely. From equation (\ref{Xray_lum}), the observed flux in X-rays from a star forming region is:
\begin{equation}
f=\frac{L}{4 \pi d_{L}^{2}}= 5 \times 10^{-20} 
\efficiency \left
( \frac{\rm SFR}{1 {\rm M_{\odot} yr^{-1}}} \right) \left( \frac{1+z}{10}
\right)^{-2} \, {\rm erg \, s^{-1} \, cm^{-2}}
\label{point_flux}
\end{equation} 
By contrast, the Chandra X-ray Observatory (CXO) sensitivity (see
http://chandra.harvard.edu/) for a 5$\sigma$ detection of a point
source in an integration time of $10^{5}$s in the 0.2--10 keV
range is $F_{X} = 4 \times  10^{-15} \, {\rm erg \, s^{-1} \,
cm^{-2}}$. The next generation X-ray telescope, Constellation-X
(http://constellation.gsfc.nasa.gov/) is optimised for X-ray
spectroscopy and will not have significantly greater point source
sensitivity: for a similar bandpass and integration time it will be
able to detect objects out to a flux limit of $F_{X} = 2 \times
10^{-15} \, {\rm erg \, s^{-1} \, cm^{-2}}$. Thus, in a week CXO and
Constellation-X will at best be able to detect starbursts (with seed photons
provided by the IR radiation field from dusty star forming regions) with star formation rates SFR $\sim 100 \,
{\rm M_{\odot} yr^{-1}}$ (corresponding to $L_{X}\sim 4.8 \times 10^{42}
\, {\rm erg \, s^{-1}}$) out to redshift $z \sim 1$ (note that the Lyman-break galaxies detected at
$z\sim 3$ (Steidel et al 1996) typically have have inferred star
formation rates ${\rm SFR} \sim 100 {\rm M_{\odot} \, yr^{-1}}$),
while the very brightest starbursts, with star formation rates of SFR$\sim 1000
\, {\rm M_{\odot} yr^{-1}}$, might be detectable out to $z
\sim 3$. As discussed in section \ref{radio_obs}, simultaneous
detection of synchrotron radiation with the Square Kilometer Array will then constrain magnetic
field strengths in these objects.

Detection in gamma ray emission is also unlikely. The upcoming
Gamma-ray Large Area Space Telescopy (GLAST)
(http://glastproject.gsfc.nasa.gov/) will have
a  flux sensitivity of
$\sim 2 \times 10^{-9} {\rm photons \, cm^{-2} \, s^{-1}}$ in the
20MeV--300GeV range, whereas star-forming regions at high redshift
will have fluxes of at most $F_{\gamma} \sim 10^{-15} \efficiency \SFR
\redshift^{-2} {\rm photons \,
cm^{-2} s^{-1}}$.

\begin{figure}
\epsscale{1.00}
\plotone{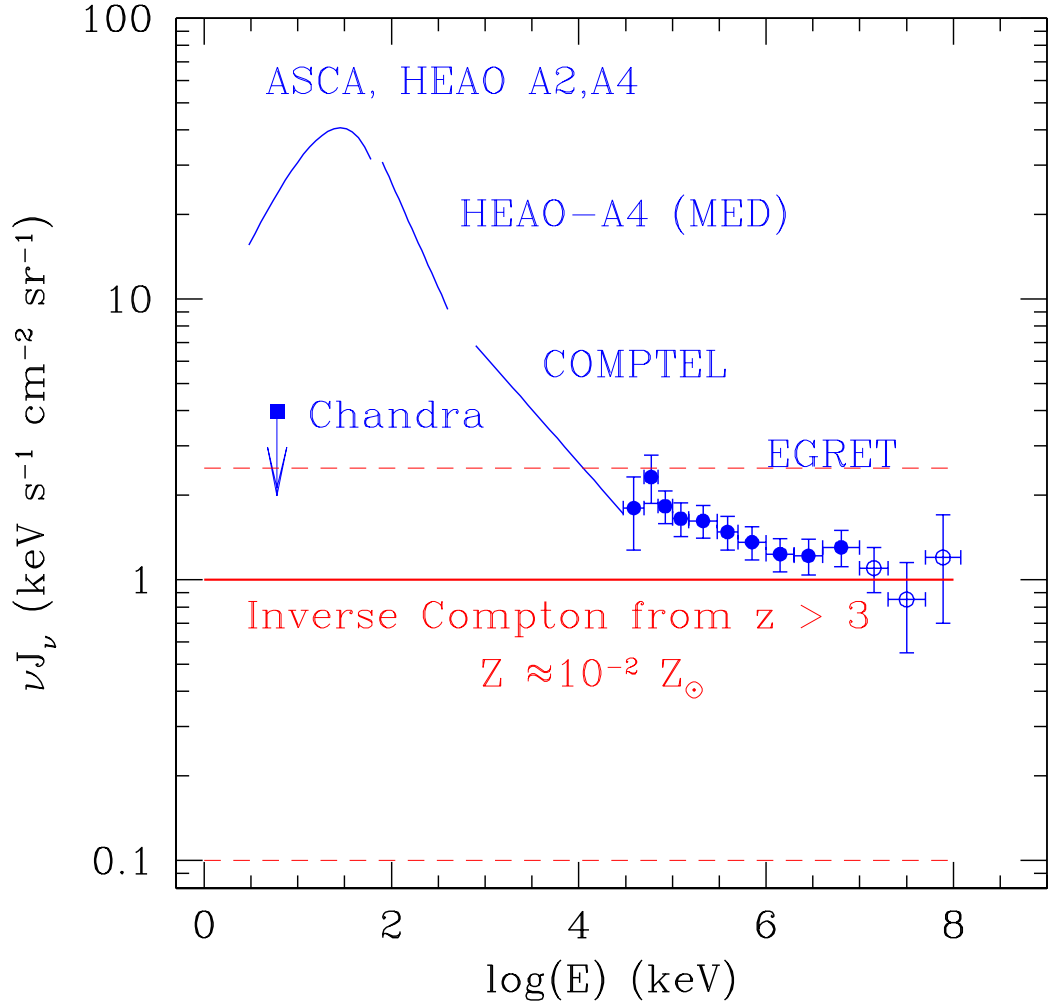}
\caption{Current best observational limits on the X-ray and gamma-ray
background, against model predictions. The solid line represents model
predictions for an IGM metallicity at z=3 of ${\rm Z \approx 10^{-2}
Z_{\odot}}$, while the upper and lower dotted lines represent generous
limits of ${\rm Z \approx 2.5 \times 10^{-2} Z_{\odot}}$ and ${\rm Z \approx 10^{-3} Z_{\odot}}$
respectively. Note that inverse Compton radiation from high redshift can account for the majority of the observed gamma-ray background.
\label{background_fig}}
\end{figure}

What fraction of the unresolved X-ray and gamma-ray background could be due to star
formation at high redshift $z>3$? Observations of
absorption in CIV and other metals in Ly$\alpha$ forest absorption lines with
$10^{14.7} {\rm cm^{-2}} < N_{HI} < 10^{16} {\rm cm^{-2}}$ indicate that $Z\sim
10^{-2} Z_{\odot} \sim 2 \times 10^{-4}$.  Each supernova produces about
$\sim 1 M_{\odot}$ of metals (Woosley \& Weaver 1995). Assuming
$(\Omega_{b},h)=(0.04,0.65)$, this implies one supernova every $\sim
1000 \, {\rm kpc^{3}}$ (comoving). If each supernova injects $\sim
10^{50}$ ergs in hard X-rays, the comoving X-ray
energy density was $U_{X} \sim \frac{c}{4 \pi} \frac{U_{X}}{(1+
\bar{z})} \sim 2 \times 10^{-6} {\rm eV \,
cm^{-3}}$. If the mean source redshift was $\bar{z} \sim 5$, the
distant sources produced a diffuse flux of $J \sim 1
{\rm keV s^{-1}
cm^{-2} sr^{-1}}$. One can make a more detailed estimate using the
equation of cosmological transfer (Peebles 1993):
\begin{equation}
J(\nu_{o},z_{o})= \frac{1}{4 \pi} \int_{z_{o}}^{\infty} dz
\frac{dl}{dz} \frac{(1+z_{o})^{3}}{(1+z)^{3}} \epsilon(\nu,z) e^{-\tau_{{\rm
eff}}(\nu_{o},z_{o},z)} 
\label{rad_transfer}
\end{equation}
where $z_{o}$ is the observer redshift, $\nu=\nu_{o}(1+z)/(1+z_{o})$,
and $\epsilon(\nu,z)$ is the comoving X-ray emissivity. Using the star formation model of Haiman
\& Loeb (1997), I obtain the estimate:
\begin{equation}
\nu J_{\nu}= 0.8 \efficiency \left( \frac{Z(z=3)}{10^{-2} Z_{\odot}}
\right) {\rm keV \, s^{-1}
cm^{-2} sr^{-1}}     \ \ {\rm for} \ z_{source} \ge 3
\end{equation}
In Fig (\ref{background_fig}), I display this predicted level of emission
against the observed X-ray and gamma-ray background in the 1 keV --100
GeV range, from analytic fits to the ASCA and HEAO A2,A4 data in the 3-60 keV range (Boldt
1987), the HEAO 1 A-4 data in the 80-400 keV range (Kinzer et al
1997), the COMPTEL data in the 800 keV -- 30MeV range (Kappadath et al
1996). Data points from EGRET in the 30 MeV -- 100 GeV
range (Sreekumar et al 1998) are also shown.  The level of the unresolved background
of course depends on the sensitivity of the instrument; recently CXO
resolved $\sim 80 \%$ of the hard X-ray background in the 2-10 keV
range into point sources (Mushotzky et al 2000). This agrees well with
the predictions of XRB synthesis models (e.g., Madau, Ghisellini \&
Fabian 1994), which use AGN unification schemes to reproduce the
observed spectral shape of the XRB. A prediction of the
IC scenario presented here is that a non-trivial fraction of the
X-ray/gamma-ray background will not be resolved into point sources
with upcoming missions, due to the extreme faintness of high redshift sources.

It is particularly intriguing that both the amplitude and spectral
shape of the gamma-ray background as observed by EGRET is well-matched
by the predicted level of gamma-ray emission in this model. This
raises the exciting possibility that the majority of the observed
gamma-ray background comes from inverse Compton emission at high
redshifts. At present, the origin of the gamma-ray background is still
unknown. The most favoured scenario for some time was that it is due
to unresolved gamma-ray blazars (Bignami et al 1979, Kazanas \&
Protheroe 1983, Stecker \& Salamon 1996): the observed blazar
$\gamma$-ray spectrum has an average spectral index compatible with
the observed GRB (Chiang \& Mukherjee 1998). However, extrapolation of
the observed EGERT blazar luminosity function implies that unresolved
blazars can account for at most $\sim 25\%$ of the diffuse
$\gamma$-ray background (Chiang \& Mukherjee 1998). The unresolved
blazar model will most likely be decisively tested by GLAST (Stecker \&
Salamon 1999), which will be two orders of magnitude more sensitive
than EGRET. A host of other models include pulsars expelled into the
halo by asymmetric supernova explosions (Dixon et al 1998, Hartmann
1995), primordial black hole evaporation (Page \& Hawking 1976),
supermassive black holes at very high redshift (Gnedin \& Ostriker
1992), annihilation of weakly interactive big bang remnants (Silk \&
Srednicki 1984, Rudaz \& Stecker 1991) and finally, inverse Compton
radiation from cosmic ray electrons in our own Galaxy (Strong \&
Moskalenko 1998, Dar \& De Rujula 2000), and from collapsing clusters
(Loeb \& Waxman 2000). However, to date the possibility of inverse
Compton emission from high redshift supernovae has not been discussed.    

A scenario in which the majority of the gamma-ray background comes
from inverse Compton emission at high redshift make a number of firm
predictions: (i) As previously mentioned, the majority of the GRB will remain unresolved by
GLAST, due to the extreme faintness of the contributing sources. (ii)
After removal of the Galactic contribution, which is correlated
with the structure of our Galaxy and our position within it (Dixon et
al 1998, Dar et al 1999), a highly isotropic component of
the GRB will still be present. (iii) The GRB should be extremely
smooth, and exhibit significant fluctuations only at extremely small
angular scales. The fluctuations should be dominated by the Poisson
rather than the clustering contribution (see Oh 1999). Using $\langle
S^{n} \rangle = \int_{0}^{S_{c}} \frac{dN}{dS} S^{n}$, where $S_{c}$ is the cut-off flux for point
source removal, and the star formation model of Haiman \& Loeb (1997), I find that for $z > 3$ sources (which are too faint
to be removed as point sources), $\frac{\langle I^{2} \rangle^{1/2}} {\langle I
\rangle}= 3.4 \times 10^{-2} \left( \frac{\theta}{5^{\prime}} \right)^{-1}$, where $I$ is surface brightness (the angular resolution of
GLAST is expected to be of order $1-5$ arcmin). (iv) The gamma-ray background at $E > 100$GeV should be attenuated, due to pair production opacity against
IR/UV photons (e.g., Salamon \& Stecker 1998, Oh 2000b). High energy
photons initiate an electromagnetic cascade which transfers energy
from high energy photons to the lower energy portion of the spectrum,
where the universe is optically thin (Coppi \& Aharonian 1997). Note that the EGRET
spectrum was directly determined with data only up to 10 GeV; beyond
10 GeV larger uncertainties exist due to backsplash in the NaI
calorimeter, and Monte-Carlo simulations were used to determine the
differential flux in the 10-30,30-50 and 50-120 GeV range (Sreekumar
et al 1998). Thus, the EGRET data points with $E > 10$GeV are less
reliable. It would be intriguing to see if GLAST (sensitive out to
300 GeV) indeed shows an absorption edge to the gamma-ray background 
at higher energies. It would also be interesting to look for
absorption of the gamma-ray background in a line of sight passing
through a massive cluster (indicating that the gamma-rays come from
higher redshifts than the cluster); the pair-production optical depth
through a cluster is of order $\tau \sim  2 n_{\gamma} \sigma_{T}
r_{vir} \sim 0.4$ (assuming $n_{\gamma} \sim 0.1 {\rm cm^{-3}}$ and
$r_{vir} \sim 1$Mpc).     

%We have only considered the effect of high redshift supernova
%explosions at $z> 3$ to the XRB and gamma-ray background. The
%contribution of low redshift galaxies to these backgrounds is more
%difficult to estimate. Dar \& De Rujula (2000) apply their inverse
%Compton model for our own Galaxy to extragalactic sources, using the
%redshift evolution of star formation given by Steidel et al (1998);
%they find at most $\sim 10\%$ of the gamma-ray background can be
%accounted for. However, this ignores
%that fact starburst galaxies reach sufficiently high seed photon energy
%densities that inverse Compton radiation becomes much more
%significant.  Thus, IC emission from
%galaxies at $ z< 3$ could be comparable to that from galaxies at $z>
%3$. For galaxies at lower redshift, the expected cutoff in the
%gamma-ray background occurs at higher energies.   
   
\subsection{CMB constraints}

Will the upscattering of CMB photons by relativistic
electrons at high redshift produce an observable signal in the CMB, or
violate any present observational constraints? When the electrons are relativistic,
CMB photons are inverse Compton scattered to such high energies
(completely out of the detector bandpass) the process may simply be
thought of as absorption, $\delta I/I \sim \tau_{e}^{rel}$, where
$\tau_{e}^{rel}$ is the optical depth of relativistic electrons, and
$I$ is the number flux of CMB photons. The flux decrement due to the absorption of CMB photons may
be estimated as:
\begin{equation}
\Delta S_{\nu} \approx J_{CMB} \int d\Omega
\int dl n_{e}^{rel} \sigma_{T}  = 
J_{CMB} \sigma_{T}
\frac{N_{e}^{rel}}{d_{A}^{2}}
\label{SZ_flux}
\end{equation}
where $J_{CMB}$ is the blackbody surface brightness of the CMB, and $N_{e}^{rel}$ is the total number of relativistic
electrons in the system. The steady state number of relativistic electrons is
given by $N_{e}^{rel} = \int d \gamma N(\gamma)$ where $N(\gamma)$ is
given by equation (\ref{electron_pop_soln}). Since the CMB flux peaks at $\nu \sim 10^{11}$Hz, the absolute magnitude of the flux
decrement is maximized by going to similar frequencies. At 20 GHz,
the highest frequency detectable by SKA, the SKA has an rms sensitivity of $\sim 6$nJy for a $10^{5}$s integration. On the other hand, the flux decrement is
$\Delta S_{\nu} (20 \, {\rm GHz}) \sim 3.5 \times 10^{-5} (\frac{1+z}{10})^{2}
(N_{e}^{rel}/3 \times 10^{58} {\rm electrons})$ nJy, which is
unobservably small. 

One might hope to detect the mean signal of all the CMB photons
upscattered by relativistic electrons. In particular, since the number
of CMB photons is no longer conserved, the absorption might be
detectable as a chemical potential distortion of the CMB. Let us
estimate the number of CMB photons destroyed. Each supernova
upscatters at most $N_{scattered} \sim \epsilon E_{SN}/\langle E_{X} \rangle \sim 10^{60}$
CMB photons, where I have set the average photon energy
$\langle E_{X} \rangle \sim 100$eV (note that since the number of photons $N_{\nu}
\propto \nu^{-1}$, most of the upscattered photons are of low
energy). For a metallicity of $Z \sim 10^{-2} Z_{\odot}$ at $z \sim
3$, one supernova
has gone off every comoving $V_{SN} \sim 1000 \, {\rm kpc^{3}}$, and thus
the comoving number density of upscattered CMB photons is $\delta n
\sim N_{scattered}/V_{SN} \sim 4 \times
10^{-8} \, {\rm cm^{3}}$. Since
$n_{\gamma} \sim 400 \, {\rm cm^{-3}}$, we have $\delta n/n \sim 10^{-10}$,
which results in an undetectably small chemical potential
distortion. Thus, the upscattering of CMB photons at high redshift
does not violate any distortion constraints on the CMB.

If the IGM is reionized inhomogeneously, as in canonical models, then
secondary CMB anistropies will be created by CMB photons Thompson
scattering off moving ionized patches (Agahanim et al 1996, Grusinov \& Hu
1998, Knox et al 1998).  The power spectrum is generally white noise,
with $\Delta T/T \sim 10^{-6}-10^{-7}$, peaking at arc-minute to sub
arc-minute scales. However, if the IGM is reionized fairly
homogeneously by X-rays then over a line of sight the positive and negative contributions
of the velocity field will cancel out. In this case, only the second-order Ostriker-Vishniac effect
due to coupling between density and velocity fields will be
present. A null detection of the inhomogeneous reionization anisotropy
could place upper limits on the patchiness of reionization, although
this will be a difficult measurement as the inhomogeneous reionization
and Ostriker-Vishniac signals are likely to be of comparable strength (Haiman \& Knox 1999). 

\subsection{Radio observations}
\label{radio_obs}

The supernovae that exploded generate magnetic turbulence and magnetic
fields, which allow Fermi acceleration to take place (Jones \&
Ellison 1991, Blandford \& Eichler 1987); observations of local radio
galaxies in non-thermal radio and X-ray emission yield field strengths
consistent with equipartition between relativistic particles and the
magnetic field (Kaneda et al 1995). If such magnetic fields are present in the first star clusters, the
relativistic electron population is a source of synchrotron radio
emission as well as inverse-Compton emission. Below I find that for a given relativistic electron population, the Square Kilometer Array (SKA) will be much more sensitive to non-thermal radio emission than
CXO or Constellation-X will be to inverse Compton X-ray emission. Radio
observations will thus allow one to establish the presence of
relativistic electrons in objects too faint to observe directly in X-ray emission. Since one
knows the CMB energy density exactly as a function of redshift, given reasonable assumptions for the magnetic field the observed radio emission allows one to immediately estimate the amount of inverse Compton X-ray emission which must be taking place, 
\begin{equation}
{\rm L_{X}}= \frac{\rm U_{CMB}(z)}{\rm U_{B}}{\rm L}_{\rm synch} 
\label{X_radio}
\end{equation}

The transition Lorentz factor $\gamma_{break}$ at which electron energy losses are
dominated by inverse Compton rather than ionization losses is given by
equation (\ref{gamma_break}). Above an observed radio
frequency $\nu_{break}= \nu_{L} \gamma_{break}^{2}/(1+z) = 2.8 \times 10^{4} \Bfield
\density \redshift^{-5}$ Hz (where $\nu_{L} \equiv \frac{eB}{2 \pi m_{e} c}$ is the electron gyrofrequency), the steady state electron population is
determined by balance between supernova injection and inverse Compton
losses, and is given by equation
(\ref{electron_pop_soln}). From standard formulae for synchrotron
emissivity (Rybicki \& Lightman 1979) $\epsilon_{\nu}^{synch}=
\sigma_{T} \gamma^{2} \beta^{2} U_{mag} c \, {\rm n(E) dE/d\nu}$ (where n(E) is
the number density of emitting electrons in $dE$) this yields a synchrotron luminosity:
\begin{equation}
L_{\nu}^{sync}= 2.7 \times 10^{28} \Bfield^{1+\alpha} \freq^{-\alpha} \SFR
\efficiency f(z,U_{\gamma}) \ {\rm erg \, s^{-1} Hz^{-1}}
\end{equation}   
where $f(z,U_{\gamma})= \radrel$ (the latter term in brackets is used if the stellar radiation
field has a higher energy density than the CMB). By contrast, the thermal free-free emission is given by (Oh 1999):
\begin{equation}
L_{\nu}^{ff}=1.2 \times 10^{27} \left( \frac{{\rm SFR}}{1 {\rm
M_{\odot} yr^{-1}}} \right) \ {\rm erg \, s^{-1} \, Hz^{-1}} 
\end{equation}
Thus, assuming $\alpha=1$, at observed frequencies $\nu < \nu_{trans} = 2.3 \redshift^{-1}
\Bfield^{2} \efficiency f(z,U_{\gamma}) \, {\rm GHz}$ 
synchrotron emission dominates
over free-free emission. This is well within the 0.1--20 GHz
capability of the SKA. However, this is only true if the power law for
electron population extends to high energies. If it is truncated at
some maximum Lorentz factor $\gamma_{max}$, then synchrotron emission
is only observable for $\nu < \nu_{max}= \nu_{L} \gamma^{2}_{max} = 0.28 \redshift^{-1} \Bfield
\left( \frac{\gamma_{max}}{10^{4}} \right)^{2} \, {\rm GHz} $. 
If $\nu_{max}< \nu_{trans}$, this will 
be manifested by an abrupt drop
in radio flux at $\nu_{max}$, beyond which the emission takes the flat
spectrum free-free emission form (this does not take place for
expected values of $\gamma_{max}= 7.4 \times 10^{6} \Bfield^{1/2} \left(
\frac{v_{sh}}{2000 \, {\rm km \, s^{-1}}} \right) \redshift^{-2}$; see
equation (\ref{Emax_eqn})). Observation of such a drop will
yield valuable constraints on $B,\gamma_{max}$ in the first
objects; the value of $\gamma_{max}$ in turn constrains the upper
energy cutoff in the inverse Compton X-ray/gamma-ray spectrum, which
could be important in determining whether high-redshift starbursts
make a significant contribution to the gamma-ray background. 

\begin{figure}
\epsscale{1.00}
\plotone{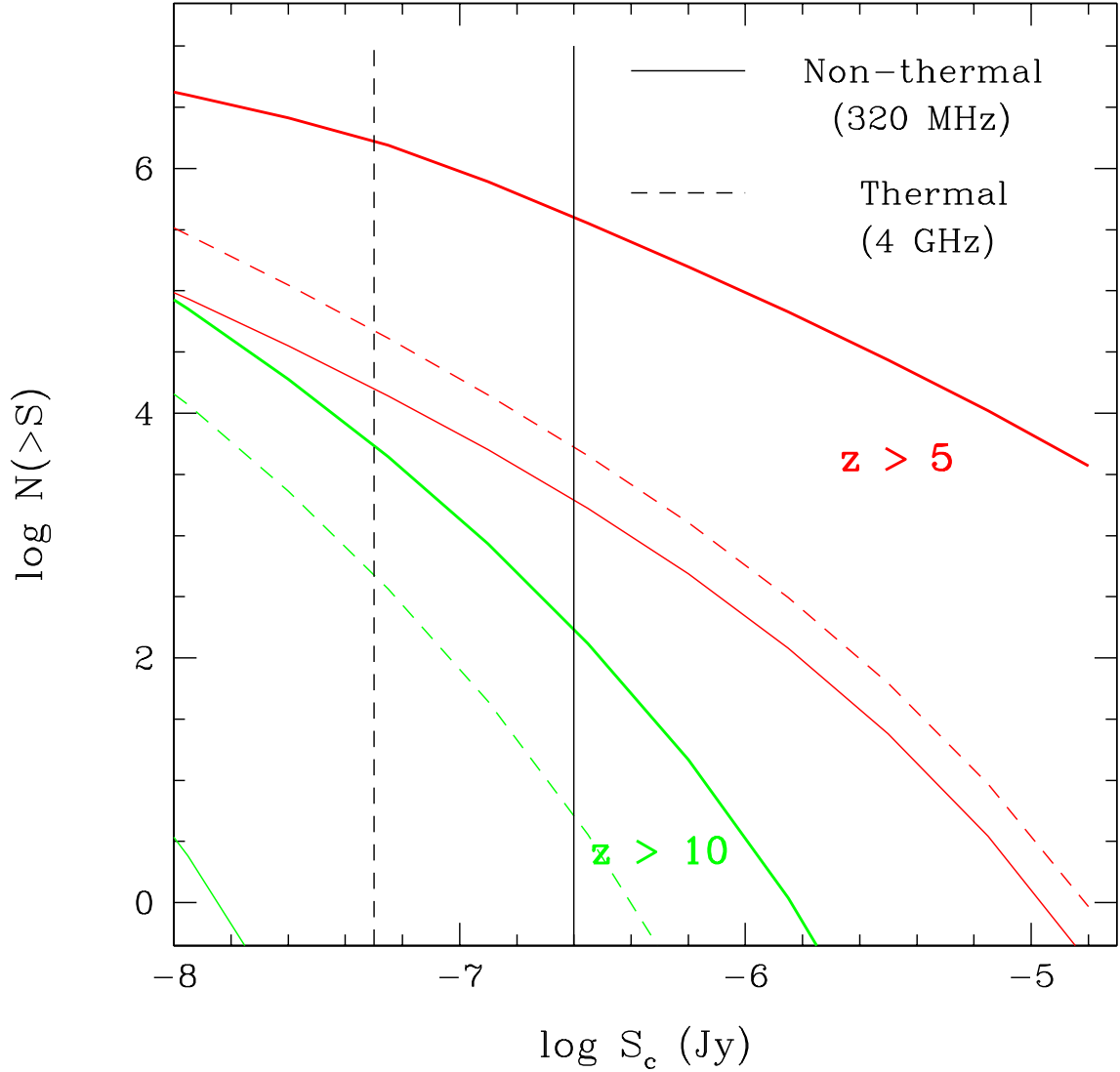}
\caption{Number of objects with $z>5,z>10$ detectable in non-thermal
emission in the $1^{\circ}$ field of view of the SKA at 320 MHz; the
limiting flux for a 10 $\sigma$ detection in a $10^{5}$s integration
is displayed as a solid line. The dark solid line is for an efficiency
factor $f_{radio} = \Bfield^{1+\alpha} \efficiency \left
( \frac{f_{star}}{0.17} \right)=1$, the light solid line is for
$f_{radio}=10^{-2}$. Also shown as dotted lines is the
detection rate and limiting flux at 4 GHz, when free-free emission is
dominant, for a high efficiency star formation model ($f_{star}=17 \%$). 
\label{synchrotron_fig}}
\end{figure}

Can radio emission from star clusters at high redshift be
detected by the proposed Square Kilometer Array? Non-thermal emission
can be distinguished from free-free emission with multi-frequency observations to identify frequency regimes where the spectral slope
is steep.The SKA detector
noise may be estimated as:
\begin{equation}
S_{\rm instrum}= \frac{2 kT_{\rm sys}}{A_{\rm eff} \sqrt{2 t \Delta \nu}}= 25
\left(\frac{\Delta \nu}{160 \, {\rm MHz}} \right)^{-1/2} \left( \frac{t}
{10^{5} \, s} \right)^{-1/2} {\rm nJy}
\label{noise}
\end{equation}
where I have used $A_{\rm eff}/T_{\rm sys}=2 \times 10^{8} {\rm
cm^{2}/K}$ for the SKA (Braun et al 1998), and I assume a bandwidth
$\Delta \nu \approx 0.5 \nu$. The flux density due to non-thermal
emission from a high-redshift star cluster, assuming $\alpha=1$, is:
\begin{equation}
S_{\nu_{o}}= \frac{L_{\nu}(\nu_{o}(1+z)}{4 \pi d_{L}^{2}} (1+z)= 8.5 \Bfield^{1+
\alpha} \left( \frac{\nu}{320 MHz} \right)^{-\alpha} \SFR
\efficiency \redshift^{-2} f(z,U_{\gamma}) {\rm nJy}
\end{equation}
Thus, a source with SFR$ \sim 10 {\rm M_{\odot} yr^{-1}}$ at $z \sim 9$ can
be detected as a 10$\sigma$ detection in 10 days. To estimate the
number of sources detectable by SKA, I use the Press-Schechter based
high redshift star formation models of Haiman \& Loeb (1997), and define an efficiency factor
$f_{radio} = \Bfield^{1+\alpha} \efficiency \left
( \frac{f_{star}}{0.17} \right)$, where $f_{star}$ is the fraction of
halo gas which fragments to form stars. In Figure
(\ref{synchrotron_fig}), I display the
number of sources above a given redshift which may be detected in
non-thermal emission in the $1^{\circ}$ SKA field of view, assuming
$f_{radio} = 1, 0.01$. Also shown is the number of sources which can
be detected in free-free emission at 4 GHz. One
should be able to detect a large number of sources at high redshift, $z>5$. Thus, radio
observations of non-thermal emission can serve as a useful proxy for
X-ray observations in allowing one to estimate $L_{X}$, and thus the
overal level of X-ray emission at high redshift.  

\subsection{Multi-wavelength observations}

While a detailed study of high-redshift multi-wavelength campaigns is beyond
the scope of this paper, below I describe some possible follow-up
observations if synchrotron emission is detected at high redshift. 
\begin{itemize}

{\item \bf Redshift estimates} Thus far the most efficient way to
select high-redshift radio galaxies has proven to be the observation
of steep radio spectra $\alpha < -1.3$ (Chambers et al 1990, van Breugel et al 1999). This is
at least partially due to the fact that extremely bright radio
galaxies have SEDs which steepen with frequency; the k-correction then
implies that sources at increasing redshift have steeper observed
spectra. Steepening with redshift due to inverse Compton losses, as
well as selection effects (i.e., brighter sources have stronger
magnetic fields, and thus more rapid synchrotron losses) could also
play a role (Krolik \& Chen 1991). This technique may fail for fainter
sources at high redshift since for $\nu > {\rm
min}(\nu_{trans},\nu_{max})$, the radio flux will be dominated by
free-free emission and the spectra will appear
flat. For instance, for $ B < 3 \redshift^{2.5} \efficiency^{0.5} \, \mu$G, we
have $\nu_{trans} < 160$MHz and SKA will only detect free-free
emission within its frequency coverage. An efficient
way to select high-redshift objects prior to reionization would be to
perform broad-band deep field imaging with NGST and  
select Lyman-break dropouts as has been done at $z \sim 3$ (Steidel et
al 1996); before the epoch of reionization one must select
'Gunn-Peterson dropouts', i.e. galaxies with no flux shortward of
rest-frame HI Ly$\alpha$. Yet another method of selecting
high-redshift objects would be to perform a joint survey in the submm
with the Atacama Large Millimeter Array (ALMA); at low redshift the
submm dust emission scales almost linearly with other star formation
indicators, such as radio and UV emission. However, for $z \approx 0.5-10$
and a given dust emission SED the K-correction almost balances the
cosmological dimming of a source, implying a flux density almost
independent of redshift (e.g., Blain et al 2000). This has spawned suggestions to obtain
approximate redshifts by the flux density ratio between submm and
radio wavebands (Carilli \& Yun 1999, 2000), although uncertainties
include an AGN contribution to the radio flux and the dust temperature
in the galaxy, with a degeneracy between hotter galaxies at high
redshift and cooler ones nearby (Blain 1999).  

{\item \bf Relative importance of X-ray and UV stellar emission for reionization} As
previously noted, from a measurement of the radio synchrotron flux we
can use equation (\ref{X_radio}) to estimate the level of inverse
Compton X-ray emission which must be taking place. Redshifts and thus
${\rm U_{CMB}(z)}$ can be determined with Balmer line spectroscopy
with NGST, while ${\rm U_{B}}$ can be estimated by assuming a
magnetic field strength required to minimise the total energy density
of the system (this is close to the value for energy equipartition
between relativistic particles and magnetic fields). Observations of
local radio galaxies in non-thermal radio and X-ray emission yield
field strengths consistent with the minimum energy value (Kaneda et al
1995). NGST can constrain the rest-frame UV emission, and a joint
measurement of the rest-frame UV flux longward of Ly$\alpha$ and the
Balmer line flux (where ${\rm L_{H\alpha}} \propto (1- f_{esc})
\dot{N}_{ion}$) can constrain $f_{esc}$, the escape fraction of
ionizing photons, where one roughly expects $J_{H\alpha} \sim 40 (1-
f_{esc}) J_{IR} (R/1000)$. Thus, joint SKA/NGST observations could place limits
on whether inverse Compton X-rays or stellar UV photons were
energetically dominant and thus more important in reionizing the universe. 

{\item \bf Measuring magnetic fields in bright sources} The very brightest sources out to
$z \sim 3$ will also be visible in X-ray emission with CXO; the fact that the same population
of relativistic electrons is responsible for non-thermal X-ray and
radio emission can be confirmed by comparing respective spectral
slopes. In this case, one can estimate the strength of magnetic
fields, ${\rm U_{B}} \approx \frac{L_{X}}{L_{sync}} {\rm
U_{CMB}(z)}$. Since dust obscuration is unimportant at both hard X-ray
and radio wavelengths, the measurement should be fairly
robust.

{\item \bf Distinguishing between synchrotron emission from AGN and supernovae}
To confirm that such emission arises from star-forming
regions rather than an AGN, one might look for signs of diffuse emission
(note that the angular resolution of the SKA is $\sim 0.1''$, while
the angular scale of the virial radius of typical objects will be
$\theta_{vir} \sim 0.5'' (M/10^{9} M_{\odot})^{1/3}$). AGNs can be
selected on the basis of color, as has been successfully carried out
at lower redshifts (Fan 1999), using broad band NIR and MIR imaging
with NGST. Finally, the line widths of the H$\alpha$, H$\beta$
lines as observed with NGST may be considerably broader for an AGN ($\sigma
\sim 1000 \, {\rm km \, s^{-1}}$), due to line broadening by the accretion disc.

\end{itemize}

\section{Conclusions}

X-ray emission from a early star forming regions is predicted to be
large and energetically comparable to UV
emission. Non-thermal
inverse Compton emission, which provides a good fit to local
observations and should become increasingly important at high
redshift, due to the evolution of the CMB energy density, is predicted
to be the dominant source of X-rays. It introduces a whole host of physical consequences:
the topology of reionization changes, becoming more homogeneous with
much fuzzier delineation between ionized and neutral regions;
reheating temperatures increase, with implications for feedback on
structure formation and the observed width of Ly$\alpha$ forest lines;
the abundance of free electrons in dense regions increases, promoting
gas phase $H_{2}$ formation, cooling, and star
formation. These effects will be considered in a companion paper (Oh
2000a). While direct detection of individual sources with CXO
or Constellation-X appears difficult, we can hope to
confirm the presence of relativistic electrons in high redshift
objects by detecting non-thermal radio emission with the Square
Kilometer Array. Given the CMB energy density at that epoch, this
yields a minimal level of inverse Compton X-ray emission. Combined
with NGST observations of rest frame UV emission, this will determine if stellar radiation or inverse Compton X-rays were the
dominant factor in reheating and reionizing the universe. In addition,
in this scenario a non-trivial fraction of the hard X-ray and gamma-ray background
comes from inverse Compton emission at high redshift. In particular,
it is possible to reproduce both the shape and amplitude of the gamma-ray background
observed by EGRET, and predict that the majority of the smooth,
isotropic gamma-ray background will remain unresolved by GLAST, which
should display attenuation above $\sim 100$GeV, from the pair
production opacity due to ambient UV/IR radiation fields.    

Many of the conclusions in this paper depend upon a scenario in which the escape fraction of UV ionizing
photons is small, $f_{esc} < 10 \%$. This assumption is well supported
by observations in the local universe, as well as theoretical
radiative transfer calculations of high redshift star forming regions, which
predict very low escape fractions, $f_{esc} \sim 0.01$ (Wood \& Loeb
1999, Ricotti \& Shull 1999, although note that the latter authors
predict a substantial escape fraction at low masses $10^{7}
M_{\odot}$). However, note
that the latter ignore gas clumping and the multi-phase structure of
the ISM. If for some reason the escape fraction is unexpectedly high
(e.g. the supernovae blow holes in the ISM through which UV photons
can escape), then the X-ray component is energetically subdominant and
stellar UV radiation dominates the reionization of the
universe. Even in this regime, the X-ray component still plays a role
in heating the gas above 15 000 K, He II reionization, and promoting gas phase
${\rm H_{2}}$ formation, as these tasks cannot be accomplished by soft
photons. The escape fraction of UV ionising photons in high redshift
objects may eventually be deduced by comparing the IR and $H\alpha$ fluxes observed by
NGST (Oh 1999).   

The most uncertain aspect of this paper is the assumed level of X-ray luminosity, $L_{X} \sim 0.1 \dot{E}_{SN}$. I
have calibrated the conversion rate via local X-ray and gamma-ray
observations of starburst galaxies and individual supernova remnants
within our galaxy, and argued that efficiency of all proposed X-ray
production mechanisms either remains constant or increases with
redshift. If the X-ray emission in local starbursts is primarily due to inverse Compton
scattering of soft IR photons (Moran \& Lehnart 1997, Moran, Lehnart
\& Helfand 1999), the empirical relation between star
formation rate and X-ray luminosity (equation (\ref{Xray_lum}))
implies an electron acceleration efficiency of $\epsilon \sim 10 \%$,
which lies at the upper limit of theoretical expectations. If instead
the empirical relation (\ref{Xray_lum}) is approximately correct but
the X-rays arise from a variety of emission mechanisms, the
importance of X-rays for reionization still holds, but specific
observational tests which rely on the inverse-Compton mechanism, such
as the gamma-ray background observations(section (\ref{xray_obs})) and
observations of radio synchrotron emission with the SKA (section
(\ref{radio_obs})) will fail.   

\section{Acknowledgements}

I am very grateful to my advisor David Spergel for his encouragement and
advice. I also thank Roger Blandford, Andrea Ferrara, Zoltan Haiman
and Ed Turner for helpful
conversations, and Bruce Draine and Michael Strauss for detailed and
helpful comments on an earlier manuscript, and the anonymous referee
for helpful comments. I thank the Institute of Theoretical Physics, Santa Barbara for its hospitality during the completion of this work. This work is supported by the NASA ATP grant NAG5-7154, and by the National Science Foundation, grant number PHY94-07194. 

%%%%\clearpage

%%%%\clearpage

\end{document}